\documentclass[a4paper]{article}
\usepackage{amsmath}
\usepackage{amsbsy}
\usepackage{amsthm}
\usepackage{amsfonts}
\usepackage{amssymb}
\usepackage{graphicx}
\usepackage{epstopdf}
\usepackage{tikz}
\usepackage{dsfont}
\usepackage{authblk}
\usepackage{hyperref}

\newtheorem{theorem}{Theorem}
\numberwithin{theorem}{section}

\numberwithin{remark}{section}

\numberwithin{equation}{section}

\begin{document}

\title{Scaling limits and universality of Ising and dimer models}

\author[1,2]{Alessandro Giuliani}
\affil[1]{\small{Universit\`a degli Studi Roma Tre, Dipartimento di Matematica e Fisica, L.go S. L. Murialdo 1, 00146 Roma, Italy}}
\affil[2]{\small{Centro Linceo Interdisciplinare {\it Beniamino Segre}, Accademia Nazionale dei Lincei, Palazzo Corsini, Via della Lungara 10,
00165 Roma, Italy.}}

\maketitle

\begin{abstract}
After having introduced the notion of universality in statistical mechanics and its importance for our comprehension of the macroscopic 
behavior of interacting systems, I review recent progress in the understanding of the scaling limit of lattice critical models, including a quantitative characterization of the limiting distribution and the robustness of the limit 
under perturbations of the microscopic Hamiltonian. Specifically, I focus on two classes of non-exactly-solvable two-dimensional systems: non-planar Ising models
and interacting dimers. In both settings, I describe the conjectures on the expected structure of the scaling limit, review the progress towards their proof, and state some of the recent 
results on the universality of the limit, which I contributed to. Finally, I outline the ideas and methods involved in the proofs, describe some of the perspectives opened by these results, and 
propose several open problems. 
\end{abstract}

\section{Universality in Statistical Mechanics: a mathematical challenge}

Statistical Mechanics (SM) aims at explaining the macroscopic behavior of matter in its different states starting from a microscopic description of the system, which involves 
an extremely large number of elementary components, such as atoms, molecules, or spins.
Due to the complexity of the microscopic structure of realistic materials and to the necessity of handling models that are accessible to theoretical 
and numerical treatments, the mathematical modelling of any system one may wish to study inevitably requires approximations and simplifications,
often quite drastic: essentially all the models studied in equilibrium and non-equilibrium statistical mechanics are `toy models', even the most challenging ones. 
As illustrative examples, think to the description of magnets in terms of Ising, XY or Heisenberg models; of disordered materials 
in terms of the Anderson model and the interacting extensions thereof (in the context of electrical conduction in the presence of lattice defects) or of the
Edwards-Anderson model (in the context of spin glasses); of anisotropic liquids in terms of monomer-dimer systems; and so on. The oversimplifications underlying the 
definitions of these models cast a dark light on the physical reliability of their predictions. A priori, there is no reason why the 
thermodynamic and correlation functions of real magnets, liquids or conducting materials should behave quantitatively (or even qualitatively) in the same way as those of 
the Ising model, the Anderson model, the dimer-monomer model, etc. 

Predictions based on these popular but oversimplified models can be reliable only if they can be shown to be robust under the choice of the microscopic Hamiltonian, that is, 
if they depend only upon general features such as symmetry, dimensionality, etc. Vaguely speaking, robustness of the macroscopic behavior of SM systems is the content of the 
\textit{universality} principle, which will be stated more precisely in two concrete mathematical settings below. For the moment, let us just observe that, in view of the previous 
considerations, this principle can be seen as \textit{the} justification for the use of toy models in the description of complex materials and, in a sense, it is what makes SM predictive and 
useful as a whole. 

Away from the critical point, where, typically, the correlations among fluctuations of local observables decay exponentially to zero at large distances, the universality of the 
behavior of the system at the macroscopic and mesoscopic level is closely related to the Law of Large Numbers and to the Central Limit Theorem (CLT) for 
weakly correlated random variables: averages of local observables converge almost surely to their expectation (the macroscopic value of the corresponding 
thermodynamic function), and their fluctuations around the mean converge, after appropriate rescaling, to normal random variables.

Things are much more subtle and interesting in the vicinity of a phase transition, where correlations among faraway fluctuations of local observables become so important that the CLT 
has no a priori reason to hold, and will in general not hold. The understanding of phase transitions is one of the central goals of SM since at least a hundred years. The existence
of several different kinds of phase transition and the characterization of the corresponding low- and high-temperature phases are among the great successes of the SM of the XX 
century. On the other hand, a complete understanding of the behavior of the system at, or close to, the critical point, is still missing, and poses several exciting challenges for 
mathematical physics and probability. Let us focus here on the case of \textit{continuous} phase transitions. 
In such a case, do fluctuations of local observables admit an interesting, non-Gaussian, mesoscopic limit? 
Is the limit \textit{robust} under a large class of perturbations of the interaction among the microscopic constituents of the system? 

These are some of the most fundamental problems of equilibrium SM since the 1960s. The theory of Wilsonian Renormalization Group (RG) \cite{W1,W2,W3}, which is among the 
greatest success of theoretical physics in the last century, was developed for quantitatively answering to these questions. It 
predicts that the `scaling limit' describing the large scale behavior of correlations at a continuous phase transition is in great generality a Euclidean Field Theory, 
which can be determined as the fixed point of an explicit semi-group (the Wilsonian RG transformation), acting on an often vaguely-defined `space of Hamiltonians'. 
The Gaussian, often dubbed `trivial', fixed points of the Wilsonian RG transformations correspond to off-critical systems or to the simplest critical ones. What about non-trivial, i.e., 
non-Gaussian, fixed points? By construction, any such fixed point turns out to be scale invariant. It has been argued that, under some reasonable additional hypotheses on the 
structure of the correlation functions and the locality of theory, such fixed points are conformally invariant \cite{P70,Po88,Z86}, i.e., described by a Euclidean Conformal Field Theory (CFT). 
If we trust this picture, we can take an axiomatic point of view: i.e., we can try to \textit{classify} the admissible non-trivial fixed point by classifying all the possible CFTs and, whenever possible, characterize their structure (by, e.g., computing their correlation functions); this task has been essentially completed in two dimensions (2D), thanks to the rich structure of the 2D conformal group (see, e.g., \cite{BPZ,KL} for the case of the `discrete series' 
of models with central charge $0<c<1$, and \cite{KRV} for the case of Liouville theory). In three dimensions (3D), this axiomatic point of view recently led to 
some spectacular developments, which allowed to compute the critical exponents associated with the non-Gaussian behavior of local observables for several non-trivial fixed point 
theories, including one that is believed to describe the scaling limit of the 3D Ising model at its critical point \cite{PRV}. 

Once that the candidate scaling limits have been constructed in this way, one is left with identifying the right one for any given class of microscopic Hamiltonians. 
While heuristically one can appeal, e.g., to symmetry considerations or to numerical constraints on the decay exponents of correlation functions 
to guess the right scaling limit for a given microscopic Hamiltonian, 
the tasks of mathematically proving that the scaling limit of the critical theory exists and it is conformal invariant, that such limit coincides with one of the candidate Euclidean CFTs, 
and that it is robust under a large class of perturbations of the microscopic interaction, are among the great challenges of modern mathematical SM. 

Even in very specific, simple, settings, many of the natural questions arising from the above premises remain open to date. However, in the last decades 
there has been remarkable progress from different viewpoints, which allowed to exhibit the first examples of conformally invariant, universal, 
scaling limits, rigorously constructed starting from lattice microscopic models. These mathematical results are mostly restricted to 2D, which is the case I will focus on from now on.
Two complementary approaches that have been, and are being, successfully used to rigorously understand universality and conformal invariance of 2D lattice SM systems are: a probabilistic one, based on random geometry,  
percolation and discrete holomorphicity; and a field theoretic one, based on constructive RG ideas. 

The first, probabilistic, method led to the complete proof of conformal invariance of the scaling limit of the 2D planar Ising \cite{CHI,CHI2,CS,DS,HJVK,HS,S10} and dimer \cite{A21,Ke00,Ke01,KOS} models. 
It has the advantage of being flexible in treating geometric
deformations of the domain and of the underlying lattice, thus leading to the first proofs of universality with respect to these kinds of
deformations. The limitation of this approach is that it is mostly restricted to exactly solved models at the `free Fermi point' (i.e., exactly solvable models, whose solution can be expressed in determinant form, 
as for Ising and dimers) and it is not flexible in dealing with perturbations of the microscopic Hamiltonian (there are a few important exceptions, notably \cite{ADTW,S01}, 
which suggest possible directions for future extensions and developments). 
The second, field theoretic, method 
led to the construction of the bulk scaling limit of several interacting, non-solvable, models, such as Ashkin-Teller and 8-vertex (8V) models \cite{BFM,GM05,Ma04}, 
interacting dimers and 6-vertex (6V) models \cite{GMT17,GMT17b,GMT20}, the sine-Gordon model on the Kosterlitz-Thouless critical line \cite{F}, 
and many others \cite{ACG13,BBS,BW,BGPS,BM,BMS,D}:
remarkably, many of these models have non-determinantal scaling limits, and the results are robust under a large class of microscopic
perturbations of the lattice Hamiltonian. Moreover, this approach led to the proof of several predictions from CFT, such as scaling relations
among critical exponents and amplitudes \cite{BFM,GMT20,BM11}, bosonization identities \cite{BW,BFM09Th}, and expression for the universal subleading contributions
to the critical free energy \cite{GM13}. A limitation of this approach is that it is restricted to ‘weakly interacting’ cases, that is, to models that
are close to a Gaussian model or to a free Fermi model. Moreover, it is not yet flexible enough for dealing with non-translationally invariant situations, including geometric perturbations of the domain
or of the underlying lattice. However, recent progress in simple domains with boundaries \cite{AGG1,AGG2} opens new perspectives for applications to general geometries and for an effective combinations of constructive RG 
ideas with probabilistic ones. 

In the following, I will review some of these advances in the specific contexts of non-planar 2D Ising models and of non-integrable perturbations of 2D dimer models, 
focusing on a selection of results obtained via the constructive RG, whose development and application to the theory of universality in 2D SM systems I contributed to. In Section \ref{sec:2}, I discuss a class of non-planar Ising models: I will first define the setting, then state the conjectures on the universality of the scaling limit at the critical point, and then, after having reviewed the known results in 
the integrable, planar, model, I will state our main results on the existence and universality of the scaling limit for the multipoint energy correlations in the plane and in the cylinder, see Theorem \ref{thm:1} and Theorem \ref{thm:2} below. In Section \ref{sec:3}, I discuss a class of non-integrable dimer models; also in this case, after having defined the setting, stated the expected structure of the scaling limit and reviewed some of the 
known results in the integrable case, I will state our main results on the fine asymptotics of the dimer-dimer correlations and on the universality of the scaling limit of the height fluctuations, see Theorem \ref{thm:3} and  
\ref{thm:4}. In Section \ref{sec:4} I will informally describe the methods of proof, and 
in Section \ref{sec:5} I will comment on perspectives and open problems.

\section{The scaling limit of non-planar Ising models}\label{sec:2}

Consider a finite, simply connected, region of the plane, $\Omega\subset \mathbb R^2$, and let $\Omega_a=\Omega\cap a\mathbb Z^2$ be its discretization 
on the square grid of lattice spacing $a>0$. At each site $x$ of $\Omega_a$ we assign an Ising spin $\sigma_{x}\in\{+,-\}$ and, given the spin configuration 
$\sigma\in\{+,-\}^{\Omega_a}$, we assume that its energy, or Hamiltonian, has the following form:
\begin{equation}H^{\lambda;\emptyset}_{a,\Omega}(\sigma)= -J\sum_{\langle x,y\rangle}\sigma_x\sigma_y-\lambda\sum_{X\subset \Omega_a} V(X)\sigma_X
,\label{1.1}\end{equation}
where: $J>0$; the first sum runs over (unordered) nearest neighbor pairs of sites in $\Omega_a$;
in the second sum, given a subset $X$ of $\Omega_a$, we denoted 
$\sigma_X:=\prod_{x\in X}\sigma_x$ and $V$ is a translationally invariant interaction, supported on \textit{even} sets $X$, of finite range proportional to $a$\footnote{More precisely, 
we assume $V\!(X)\!=\!V_0(X/a)$ for a fixed, finite range, potential $V_0:\mathbb Z^2\to\mathbb R$.}.
In general, we will require $V$ neither to be a pair interaction (i.e., to be supported on sets of cardinality $2$) nor to be ferromagnetic (i.e., to be non-negative).
We will refer to the first term in the right side of \eqref{1.1} as to the nearest-neighbor interaction, of strength $J$ (to be fixed once and for all), and to the second term as to the multi-spin 
interaction, of strength $\lambda$ (to be thought of as being small as compared to $J$). The model describes the magnetic properties of thin ferromagnetic films with an out-of-plane easy-axis of magnetization and short range `exchange' interactions among the magnetic moments of the ions. For $\lambda=0$ the Hamiltonian reduces to that of the 
planar, nearest-neighbor, Ising model, originally introduced by Lenz in 1920 \cite{L20}, which in 2D is exactly solvable in a very strong sense, as originally proved by Onsager \cite{O44}, 
see Section \ref{sec:2.1} below. For $\lambda\neq 0$, the multispin interaction breaks planarity 
(i.e., the interaction cannot be represented in terms of couplings associated with the edges of a planar graph with edge set $\Omega_a$), as well as the integrability of the model. 

The apex $\emptyset$ on $H^{\lambda;\emptyset}_{a,\Omega}$ refers to the boundary conditions (b.c.), which we implicitly assumed to be `open', or `free', i.e., we assumed that there is no 
spin in the complement of $\Omega_a$ interacting with those in $\Omega_a$. 
In a similar way, we can define the Hamiltonians $H^{\lambda;+}_{a,\Omega}$ (resp. $H^{\lambda;-}_{a,\Omega}$) with $+$ (resp. $-$) b.c., 
by including in its definition the interactions between the spins $\sigma$ in $\Omega_a$ and a configuration of spins identically equal to $+1$ (resp. $-1$) in its complement, $\Omega_a^c$.
Analogously, if $\Omega$ is a 2D torus (resp. a 2D cylinder), we denote by $\Omega_a$ its discretization of lattice spacing $a>0$ and let 
$H^{\lambda;\text{per}}_{a,\Omega}$ (resp. $H^{\lambda;\text{cyl}}_{a,\Omega}$) be the spin Hamiltonian defined as in \eqref{1.1}, with the first sum including the nearest neighbor pairs 
winding up over the torus (resp. cylinder) and the interaction $V$ being translationally invariant with respect to the natural translations on the torus (resp. cylinder). 

The finite volume Gibbs measure with inverse temperature $\beta>0$ and $\#$ b.c., with $\#\in\{\emptyset,+,-,\text{per},\text{cyl}\}$, is characterized by the probability weight
\begin{equation} \mathbb P^{\lambda;\#}_{\beta;a,\Omega}(\sigma)=\frac1{Z^{\lambda;\#}_{a,\Omega}}e^{-\beta H^{\lambda;\#}_{a,\Omega}(\sigma)}, \quad \forall \sigma\in\{+,-\}^{\Omega_a}, 
\label{eq:gibbsIs}\end{equation}
where $Z^{\lambda;\#}_{a,\Omega}=\sum_{\sigma\in\{+,-\}^{\Omega_a}}e^{-\beta H^{\lambda;\#}_{a,\Omega}(\sigma)}$ is the partition function. Given an observable 
$A:\{+,-\}^{\Omega_a}\to \mathbb R$, we denote its average with respect to (w.r.t.) the probability weight \eqref{eq:gibbsIs} by $\mathbb E^{\lambda;\#}_{\beta;a,\Omega}(A)$. 
`Truncated', or `connected', expectations are denoted by semicolons: e.g., $\mathds E^{\lambda;\#}_{a,\Omega}(A_1;A_2)=\mathds E^{\lambda;\#}_{a,\Omega}(A_1\,A_2)-\mathds E^{\lambda;\#}_{a,\Omega}(A_1)\mathds E^{\lambda;\#}_{a,\Omega}(A_2)$. 

It is well known that the system displays a phase transition, in the following sense. Fix $a>0$ and take $\lambda$ sufficiently small compared to $J$. Denote 
by $\Omega\nearrow \mathbb R^2$ the `thermodynamic limit' obtained by (say) centering 
$\Omega$ at the origin, rescaling its linear dimensions by $L$, and letting $L\to\infty$. Then:\\
$\bullet$ If $\beta$ is small enough, for any finite $X\subset a\mathbb Z^2$, 
the limit $\mathbb E^\lambda_{\beta;a,\mathbb R^2}(\sigma_X):=\lim_{\Omega\nearrow \mathbb R^2}\mathbb E^{\lambda;\#}_{\beta;a,\Omega}(\sigma_X)$ is independent of the b.c. $\#$, 
it is translationally invariant and characterized by the fact that $\mathbb E^\lambda_{\beta;a,\mathbb R^2}(\sigma_x)=0$, $\forall x\in a\mathbb Z^2$, and that the truncated correlations $\mathbb E^\lambda_{\beta;a,\mathbb R^2}(\sigma_X;\sigma_Y)$ 
decay exponentially to zero as the distance between the finite sets $X, Y\subset a\mathbb Z^2$ diverges. \\
$\bullet$ If $\beta$ is large enough, for any finite $X\subset a\mathbb Z^2$, the limits $\mathbb E^{\lambda,\pm}_{\beta;a,\mathbb R^2}(\sigma_X):=\lim_{\Omega\nearrow \mathbb R^2}\mathbb E^{\lambda;\pm}_{\beta;a,\Omega}(\sigma_X)$ with b.c. $+$ or $-$ exist, they are translationally invariant, but are different if $|X|$ is odd: in particular, $\mathbb E^{\lambda;+}_{\beta;a,\mathbb R^2}(\sigma_x)=-\mathbb E^{\lambda;-}_{\beta;a,\mathbb R^2}(\sigma_x)$ is positive and independent of $x$, and $\mathbb E^{\lambda;\pm}_{\beta;a,\mathbb R^2}(\sigma_X;\sigma_Y)$
decay exponentially to zero as the distance between the finite sets $X, Y\subset a\mathbb Z^2$ diverges. 

The two scenarios described in these items are usually referred to as `high-tempera- -ture' and `low-temperature' phases, respectively. If $\lambda V$ is a ferromagnetic pair interaction, 
it is known \cite{ABF87} that they extend to two contiguous intervals $(0,\beta_c)$ and $(\beta_c,\infty)$ separated by a 
critical inverse temperature $\beta_c=\beta_c(\lambda)$, at which, for any finite $X\subset a\mathbb Z^2$, the limit 
$\mathbb E^\lambda_{\beta_c;a,\mathbb R^2}(\sigma_X):=\lim_{\Omega\nearrow \mathbb R^2}\mathbb E^{\lambda;\#}_{\beta_c;a,\Omega}(\sigma_X)$ is independent of $\#$, 
and $\mathbb E^\lambda_{\beta_c;a,\mathbb R^2}(\sigma_x)=0$, $\forall x\in a\mathbb Z^2$. The same is expected to hold for a general (translationally invariant, even, of finite range $\propto a$) interaction $V$, 
provided that $\lambda$ is small enough.
Moreover, it is expected that $\mathbb E^{\lambda}_{\beta_c;a,\mathbb R^2}(\sigma_X;\sigma_Y)$ 
decays \textit{algebraically} to zero as the distance among the finite sets $X,Y\subset a\mathbb Z^2$ diverges. Even more, the scaling limit of the correlations is expected to exist and 
to be universal, in the following sense. Fix $\beta=\beta_c$, and let $\Omega$ be a prescribed subset of the plane, or a 2D torus, or a 2D cylinder. Define the rescaled spin and energy 
variables as:
\begin{equation}\sigma(x)=a^{-1/8}\sigma_{[x]}, \quad \epsilon_j(x)=a^{-1}\big(\sigma_{[x]}\sigma_{[x]+a\hat e_j}-\mathbb E^{\lambda;\#}_{\beta_c;a,\Omega}(\sigma_{[x]}\sigma_{[x]+a\hat e_j})\big),
\end{equation}
where, for $x\in \Omega$, $[x]=a\lfloor a^{-1}x\rfloor$, $\hat e_j$ is the unit coordinate vector in direction $j\in\{1,2\}$, and $\#\in\{\emptyset,+,-,\text{per},\text{cyl}\}$. Then it is expected that, for any tuple of distinct 
points $x_1,\ldots,x_n, y_1$, $\ldots$, $y_m$ of $\Omega$ and any choice of $j_1,\ldots,j_m\in\{1,2\}$, the limit 
\begin{equation} \lim_{a\to 0}\mathbb E_{\beta_c;a,\Omega}^{\lambda;\#} \big(\sigma(x_1)\cdots \sigma(x_n)\epsilon_{j_1}(y_1)\cdots\epsilon_{j_m}(y_m)\big)\label{2.4}\end{equation} 
exists, it is conformally covariant under Riemann mappings of the domain $\Omega$ into an arbitrary new domain $\Omega'$
and, moreover, it depends on $\lambda$ in an extremely simple, multiplicative, way: i.e., one expects that there exist two constants $Z_1=Z_1(\lambda)$ and $Z_2=Z_2(\lambda)$
such that the limit in \eqref{2.4} equals $Z_1^nZ_2^m$ times the limit obtained in the nearest neighbor case $\lambda=0$. This is what universality predicts in this context and whose proof represents 
a key challenge in mathematical SM for the incoming years. As anticipated in the introduction, lately there has been remarkable progress towards its proof, as reviewed in the next subsections. 

\subsection{The nearest neighbor case}\label{sec:2.1}

As mentioned above, for $\lambda=0$ the model is exactly solvable in a remarkably strong sense \cite{HG, KW, KaCe, KO, McCWbook, McPeW, MPW, O44, SLM, Ya52}: in particular, the partition function can be written as the Pfaffian of a suitable complex adjacency matrix $K$ of a graph, known as the Fisher graph, obtained by suitably decorating the one associated with $\Omega_a$ \cite{Fi66,Ka}; moreover, correlation functions of local observables can be expressed in terms of Pfaffians of sub-matrices of $K$. The exact solution provides, among other things, closed formulas for the free energy, specific heat, magnetization, and the large distance asymptotics of the spin correlations. 
The critical temperature is known to be $\beta_c=\beta_c(0)=(2J)^{-1}\log(\sqrt2+1)$, at which, letting first $\Omega\nearrow \mathbb R^2$ and then $a\to 0$, one finds, for any pair of distinct points $x_1,x_2\in\mathbb R^2$
and any choice of $j_1,j_2\in\{1,2\}$:
\begin{equation}
\begin{split} & \lim_{a\to 0}\lim_{\Omega\nearrow \mathbb R^2} \mathbb E_{\beta_c;a,\Omega}^{0;\#}\big(\sigma(x_1)\sigma(x_2)\big)=\frac{A}{|x-y|^{1/4}},\\
& \lim_{a\to 0}\lim_{\Omega\nearrow \mathbb R^2}\mathbb E_{\beta_c;a,\Omega}^{0;\#}\big(\epsilon_{j_1}(x_1);\epsilon_{j_2}(x_2)\big)=\frac1{\pi^2}\frac1{|x-y|^{2}},\end{split}
\end{equation}
irrespective of the boundary conditions (in the first line, $A=0.70338016\cdots$). The exact solution, 
in the form reviewed e.g. in \cite{McCWbook}, also allows one to compute the multipoint energy correlations in the infinite plane limit:
for any $n$-ple of distinct points $x_1,\ldots,x_n$ and any $j_1,\ldots,j_n\in\{1,2\}$, we have
\begin{equation}
 \lim_{a\to 0}\lim_{\Omega\nearrow \mathbb R^2}\mathbb E_{\beta_c;a,\Omega}^{0;\#}\big(\epsilon_{j_1}(x_1)\cdots\epsilon_{j_n}(x_n)\big)=\pi^{-n}\big|\text{Pf}\, M(z_1,\ldots,z_n)\big|^2,
 \label{2.6}\end{equation}
irrespective of the boundary conditions, where $z_j=(x_j)_1+i(x_j)_2$ is the complex representative of the point $x_j$, and $M(z_1,\ldots,z_n)$ is the $n\times n$ anti-symmetric matrix of elements $M_{ij}(z_1,\ldots,z_n)=\frac{\mathds{1}_{i\neq j}}{z_i-z_j}$. 
While these results are classical, conformal covariance of the limits in finite domain remained elusive for decades. In the case of the multipoint energy correlations, a rigorous proof is due to Hongler \cite{Ho_thesis}, who proved that, for any open, simply connected region $\Omega$ of the plane, letting $\varphi:\Omega\to\mathbb H$ be the conformal mapping from $\Omega$ to the upper half-plane (both thought of as subsets of $\mathbb C$), and defining $\mathbb E^{0;\#}_{\beta_c;\Omega}(\epsilon(z_1)\cdots \epsilon(z_n)):=\lim_{a\to 0}\mathbb E^{0;\#}_{\beta_c;a,\Omega}(\epsilon_{j_1}(x_1)\cdots\epsilon_{j_n}(x_n))$ for any $n$-ple $x_1,\ldots,x_n\in\Omega$ and $j_1,\ldots,j_n\in\{1,2\}$ (here, as above, $z_j$ is the complex representative of $x_j$), then, for $\#\in\{\emptyset,+,-\}$, 
\begin{equation}\label{eq:conf_en}\mathbb E^{0;\#}_{\beta_c;\Omega}(\epsilon(z_1)\cdots \epsilon(z_n))=\Big(\prod_{i=1}^n|\varphi'(z_i)|\Big)\mathbb E^{0;\#}_{\beta_c;\mathbb H}(\epsilon(\varphi(z_1))\cdots \epsilon(\varphi(z_n))).\end{equation}
Moreover, the right side is explicit: in fact, $\mathbb E^{0;\emptyset}_{\beta_c;\mathbb H}(\epsilon(z_1)\cdots \epsilon(z_n))=(-1)^n\mathbb E^{0;+}_{\beta_c;\mathbb H}$ $(\epsilon(z_1)\cdots\epsilon(z_n))$, and $\mathbb E^{0;\pm}_{\beta_c;\mathbb H}(\epsilon(z_1)\cdots \epsilon(z_n))=(i\pi)^{-n} \text{Pf}\, M(z_1,\ldots, z_n,\overline{z_n}, \ldots, \overline{z_1})$. 

The scaling limit of the multipoint spin correlations is more subtle. Kadanoff \cite{Ka69} first guessed its expression in the special case of $n$ co-linear points. In the general case, the result was
conjectured on the basis of CFT methods: in fact, after the work of Belavin, Polyakov and Zamolodchikov \cite{BPZ}, it became clear that the scaling limit of any mixed multipoint spin-energy correlation 
should coincide with the corresponding correlations of the CFT minimal model with central charge $c=1/2$, which can be explicitly computed via Coulomb Gas methods. A rigorous proof of the 
validity of the expected formula is very recent, compared with the history of the Ising model, and is due to Dubedat \cite{Du11,Du15} and to Chelkak, Hongler and Izyurov \cite{CHI,CHI2}, who proved that for even $n$, letting again $z_j$ 
be the complex representative of $x_j$, 
\begin{equation}
\lim_{a\to 0}\lim_{\Omega\nearrow \mathbb R^2}\mathbb E_{\beta_c;a,\Omega}^{0;\#}\big(\sigma(x_1)\cdots\sigma(x_n)\big)
=\Big[\Big(\frac{A}{\sqrt2}\Big)^n\sum_{\substack{\mu_1,\ldots,\mu_n=\pm:\\ \mu_1+\cdots+\mu_n=0}}
\prod_{1\le i<j\le n}|z_i-z_j|^{\mu_i\mu_j/2}\Big]^{1/2}.
\end{equation}
Even more, \cite{CHI} proved that in a finite domain $\Omega$ with $\#\in\{\emptyset,+,-\}$ boundary conditions, the limiting spin correlations, 
$\mathbb E^{0;\#}_{\beta_c;\Omega}(\sigma(z_1)\cdots \sigma(z_n)):=\lim_{a\to 0}\mathbb E^{0;\#}_{\beta_c;a,\Omega}$ $(\sigma(x_1)\cdots\sigma(x_n))$ are conformally covariant in a sense analogous to \eqref{eq:conf_en}, i.e.,
\begin{equation}\label{eq:conf_sp}\mathbb E^{0;\#}_{\beta_c;\Omega}(\sigma(z_1)\cdots \sigma(z_n))=\Big(\prod_{i=1}^n|\varphi'(z_i)|^{1/8}\Big)\mathbb E^{0;\#}_{\beta_c;\mathbb H}(\sigma(\varphi(z_1))\cdots\cdots\sigma(\varphi(z_n))),\end{equation}
and, again, the right side is explicit, see \cite[eq.(1.2)]{CHI}. 

\subsection{The non-planar case}

The remarkable results reviewed in the previous subsection are crucially based on the underlying exact solvability and discrete holomorphicity of the model.
For $\lambda\neq0$, neither of these properties holds, and completely different methods must be employed for constructing the scaling limit. As explained in the introduction,
the natural framework for treating the effect of interactions are multiscale methods, rigorously implementing Wilson's RG ideas in the present context. A constructive approach based on these ideas
was proposed in \cite{Sp,PiSp}, and successfully used in \cite{BFM,GM05, Ma04} to compute the large distance asymptotics of correlation functions and prove several instances of universality in spin and vertex models such as Ashkin-Teller, the 8V model and non-integrable variants thereof. See \cite{MaICM} for a review of these developments 2010. 
In the context of non-planar Ising models, a decade ago we successfully employed these methods to compute and prove universality of the bulk energy correlations, as sum: 

\begin{theorem}[\cite{GGM12}] Fix a potential $V$ that, besides being even and translationally invariant, has finite range proportional to $a$ and is invariant under discrete rotations and reflections. Then there exists $\lambda_0>0$ and two functions $\beta_c=\beta_c(\lambda)$ and $Z_2(\lambda)$, real-analytic in $\lambda$ for $|\lambda|\le \lambda_0$, such that, if $\Omega=\Omega_L$ is a 2D torus of side $L$, for any $n>1$, any $n$-ple of distinct points $x_1,\ldots,x_n\in\mathbb R^2$, and any choice of $j_1,\ldots,j_n\in\{1,2\}$, 
\begin{equation}\lim_{a\to 0}\lim_{L\to\infty} \mathbb E^{\lambda;\text{per}}_{\beta_c;a,\Omega}\big(\epsilon_{j_1}(x_1)\cdots \epsilon_{j_n}(x_n)\big)= 
Z_2^n \pi^{-n}\big |\textrm{Pf}\ M(z_1,\ldots,z_n)\big|^2,\label{2.8}\end{equation}
where $M(z_1,\ldots,z_n)$ is the same defined after \eqref{2.6}.\label{thm:1} \end{theorem}

Note that the right side of \eqref{2.8} is equal to $Z_2^n$ times the bulk scaling limit of the nearest neighbor model \eqref{2.6}: therefore, this theorem proves the universality conjecture for the full-plane
multipoint energy correlations of a large class of perturbations of the standard 2D Ising model. The proof of the theorem, which is based on multiscale cluster expansion methods (see Section \ref{sec:4} below)
gives many more informations than those summarized in its statement: for instance, it provides an explicit, and essentially optimal, speed of convergence to the limit, as well as 
a constructive algorithm for computing $\beta_c(\lambda)$ and $Z_2(\lambda)$; moreover, it can be adapted to 
more general cases: e.g., it does requires neither that $V$ is invariant under discrete rotations and reflections (the assumption has just the effect of simplifying the explicit expression in the right side of \eqref{2.8}), 
nor that $\beta$ is fixed exactly at $\beta_c$: choosing $\beta=\beta_c+a m_0$, we prove in \cite{GGM12} that the scaling limit of the truncated energy correlations is non-trivial (it realizes the so-called `massive scaling limit' in the temperature direction)
and it decays exponentially to zero at large distances, with rate proportional to $m_0$. 

Theorem \ref{thm:1} and its proof are restricted to a translational invariant setting, which guarantees, in particular, that the effective potentials used to describe the system at length scales much larger than the lattice spacing, 
see Section \ref{sec:4} below for details,  
can be parametri- -zed by a finite number of `relevant' and `marginal' scale-dependent couplings (using a terminology borrowed from the Wilsonian RG jargon), which are in fact \textit{constants}, rather than functions 
of the position $x$ in the domain where the system is defined on. If we are interested in constructing the scaling limit of the  correlations in a finite domain $\Omega$, then 
we need to keep track of such $x$-dependence, and to control the boundedness of the relevant and marginal couplings, as the length scale increases, uniformly in $x$. 
From a technical point of view, the $x$-dependence of the scale-dependent couplings potentially induces additional logarithmic divergences in the theory, 
arising from the integration of the degrees of freedom supported in the vicinity of the boundary. To date, there are no systematic, well-developed, methods for dealing with these divergences and related technical issues: 
the multiscale cluster expansion which the proof of Theorem \ref{thm:1} is based on is not well developed yet in the case of critical theories in finite domains, where boundaries are present and affect the 
form of correlation functions in the scaling limit. This is a severe limitation for the rigorous construction of scaling limits in finite domains and for the study 
of their conformal covariance with respect to deformations of the domain. Recently, we managed to overcome several of these technical issues and to provide the first construction of 
non-planar Ising models in a domain with boundary, in cylindrical geometry: 

\begin{theorem}[\cite{AGG1,AGG2}] Fix $V$ as in Theorem \ref{thm:1} and let $\lambda_0, \beta_c=\beta_c(\lambda)$ and $Z_2=Z_2(\lambda)$ be the same introduced there. Let $\Omega$ be a 2D cylinder
with arbitrary sides $\ell_1,\ell_2>0$, periodic in the horizontal direction. Then, for any $n>1$, any $n$-ple of distinct points $x_1,\ldots,x_n\in\mathbb R^2$, and any choice of $j_1,\ldots,j_n\in\{1,2\}$, 
\begin{equation}\begin{split}\lim_{a\to 0}\mathbb E^{\lambda;\text{cyl}}_{\beta_c;a,\Omega}\big(\epsilon_{j_1}(x_1)\cdots \epsilon_{j_n}(x_n)\big)&= 
Z_2^n \lim_{a\to 0}\mathbb E^{0;\text{cyl}}_{\beta_c;a,\Omega}\big(\epsilon_{j_1}(x_1)\cdots \epsilon_{j_n}(x_n)\big)\\
&=Z_2^n(-\pi)^{-n}\textrm{Pf}\ A(x_1,\ldots,x_n).\end{split}\label{2.9}\end{equation}
Here $A(x_1,\ldots,x_n)$ is the $2n\times 2n$ antisymmetric matrix with elements $A_{(i,a),(j,b)}$ $(x_1,\ldots,x_n)=\mathds 1_{i\neq j}g_{a,b}^{\text{cyl}}(x_i,x_j)$, where: $i,j\in\{1,\ldots,n\}$, $a,b\in\{1,2\}$,  
\begin{equation} g^{\text{cyl}}_{a,b}(x,y)= \sum_{n\in\mathbb Z^2}(-1)^n\big[g_{a,b}(x-y+\ell_n)+(-1)^a g_{a,b}(x-\tilde y+\ell_n)\big],\end{equation}
$g_{a,b}$ are the matrix elements of $g(x)=|x|^{-2}\begin{pmatrix} x_1 & x_2 \\ x_2 & -x_1\end{pmatrix}$, $\ell_n=(n_1\ell_1,n_2\ell_2)$ and $\tilde y=(y_1,-y_2)$. \label{thm:2}
\end{theorem}

Also in this case, as for Theorem \ref{thm:1}, the proof of the theorem provides bounds on the speed of convergence to the limit, and it does not rely on the assumptions 
that $V$ is invariant under discrete rotations and reflections, and that the inverse temperature is fixed exactly at $\beta_c$ (these were made just to simplify the statement). 
The key new ingredients in the proof, compared with the one of Theorem \ref{thm:1}, are the following (I use again the Wilsonian RG jargon, for additional details see Section \ref{sec:4} below): (1) proof that the scaling dimension of boundary operators is better by one dimension than their bulk counterparts, 
(2) a cancellation mechanism based on an approximate image rule for the fermionic two-point function allows us to control the RG flow of the marginal boundary terms. I expect that these novel ingredients will play an
important role in future developments in the mathematical construction of the scaling limit of critical 2D SM models in domains with boundaries. 

Let me emphasize that the result summarized in Theorem \ref{thm:2} is uniform 
in $\ell_1,\ell_2$\footnote{Strictly speaking, the proof in \cite{AGG1} requires $\ell_1/\ell_2$ to be bounded from above and below. This limitation can be easily overcome: if 
$\ell_1\ll\ell_2$ or $\ell_1\gg \ell_2$, one needs to separately study the contributions from the intermediate length scales between $\ell_1$ and $\ell_2$, 
which is easy to do by the multiscale methods of \cite{AGG1}: in fact, at these scales, the systems effectively behaves as a 1D Ising system (whose thermodynamic behavior is `trivial')
with dressed parameters.}.
Letting $\ell_1, \ell_2\to\infty$, we obtain the correlations in the half-plane (or, if desired, those in the full-plane, depending on the way in which we perform the limit). The proof can be generalized to the computation of 
the scaling limit of the boundary spin correlations, which can be shown to be the Pfaffian of an explicit anti-symmetric matrix. 
A limitation of our result, intrinsic in the multiscale cluster expansion method employed, is the restriction to small values of $\lambda$. 
A related result \cite{ADTW} is the recent proof that, if $V$ is a ferromagnetic pair interaction, then the scaling limit of the boundary spin correlations has a Pfaffian structure. The proof is 
based on a random current representation and a multiscale application of Russo-Seymour-Welsh type bounds \cite{Ru,SW} on the crossing probabilities of the currents, and applies to 
ferromagnetic pair interaction of \textit{any strength}; i.e., remarkably, the result is non-perturbative. A limitation is that the scaling limit of the boundary spin correlations constructed in \cite{ADTW}
and the associated critical exponents are not explicit: in this sense, \cite{ADTW} offers a complementary perspective to ours, both w.r.t the results and of the techniques employed. 

\section{The scaling limit of interacting dimer models}\label{sec:3}

Let us now consider a different setting, the one of 2D dimer models, where the notion of universality and the nature of the scaling limit is more subtle than for Ising models. 
Dimer models (possibly in the presence of vacancies, called `monomers') were introduced in the equilibrium setting by Fowler and Rushbrooke in 1935 \cite{FR35}
as simplified models for liquids of anisotropic molecules. Here we consider 
such models in the limit of no vacancies. Let us define the setting precisely: similarly to the previous section, let $\Omega$ be a simply connected 
region of the plane, or a 2D torus, or a 2D cylinder. Let $\mathbb L$ be the infinite square grid of lattice spacing $1/\sqrt2$ and axes tilted by $45^o$ w.r.t. the standard horizontal and vertical axes, 
and let $\Omega_a$ be a discretization
of $\Omega$ on $a\mathbb L$, $a>0$. Note that the graph $G_{\Omega_a}$ associated with $\Omega_a$, i.e., the one with vertex set $\Omega_a$ and edge set consisting of the
links connecting nearest neighbor sites of $\Omega_a$, is bipartite, and we color its vertices black and white
so that neighboring vertices have different colors, with the convention that the origin is black. 
An edge $e$ of $G_{\Omega_a}$ is said to be of type $r\in\{1,2,3,4\}$ if its white endpoint is to the NE, NW, SW, SE of its black endpoint, respectively. 
For any $e$, we let $r(e)$ be its type and $x(e)$ the coordinate of its black site (note that $x(e)\in a\mathbb Z^2$). 
 A dimer covering, or `allowed dimer configuration', of the graph $G_{\Omega_a}$ is a subset of its edges that covers every vertex exactly once. 
We denote by $\mathcal D=\mathcal D(\Omega_a)$ the set of allowed dimer configurations in $\Omega_a$, and we assume that $\Omega_a$ has been chosen in such a way that $\mathcal D\neq\emptyset$. 
The class of dimer models we are interested in are defined by the following probability measure on $\mathcal D$:
\begin{equation} \mathbb P^\lambda_{a,\Omega}(D)=\frac1{Z^\lambda_{a,\Omega}}\Big(\prod_{e\in D} t_{r(e)}\Big)e^{\lambda V(D)},\quad \forall D\in\mathcal D,\label{3.1}\end{equation}
where: $t_r$, with $r\in\{1,2,3,4\}$, are the weights of the 4 different dimer types; $V$ is a translationally invariant interaction, of finite range proportional to $a$;
$\lambda$ is the interaction strength, to be thought of as `small'; $Z^\lambda_{a,\Omega}=\sum_{D\in\mathcal D}\Big(\prod_{e\in D} t_{r(e)}\Big)$ $e^{\lambda V(D)}$ is the partition function.
With no loss of generality we can fix $t_4=1$, and we shall do so in the following. 
For $\lambda=0$ the model is exactly solvable in a very strong sense, as originally proved by Kasteleyn \cite{Ka} and by Temperley and Fisher \cite{TF}, see Section \ref{sec:3.1} below.
In general, for $\lambda\neq0$, the model is not exactly solvable anymore, even though there is a special choice of the interaction $V$, of nearest neighbor type, for which it 
reduces to the 6V model, which is solvable by Bethe ansatz, see \cite[Section 2.3]{GMT20} and \cite{Ba}. 

In analogy with the notation of the previous section, we denote by $\mathds E^{\lambda}_{a,\Omega}(A)$ the average of an observable $A:\mathcal D\to\mathbb R$ w.r.t. the probability weight in \eqref{3.1}; again, 
truncated expectations are denoted by semicolons. 
Given an edge $e$, we denote by $\mathds 1_e$ the corresponding `dimer observable', i.e., the characteristic function of the event `$e$ belongs to the dimer configuration'; (truncated)
expectations of products of dimer observables will be referred to as (truncated) dimer correlations. 
Another important observable is the height function $h$, which is defined on the faces of $G_{\Omega_a}$ as follows: fix arbitrarily a face $\eta_0$ of $G_{\Omega_a}$, 
and set $h(\eta_0)=0$; the value of the height on the other faces is fixed by letting the gradients be 
\begin{equation} h(\eta')-h(\eta)=\sum_{e\in C_{\eta\to\eta'}}\sigma_e (\mathds 1_e-1/4),\label{3.2} \end{equation}
where $C_{\eta\to\eta'}$ is a nearest neighbor path on the dual of $G_{\Omega_a}$ from the face $\eta$ to the face $\eta'$, the sum is over the edges crossed by this path, and $\sigma_e$ is a sign, equal to $+$ or $-$ 
depending on whether the oriented path $C_{\eta\to\eta'}$ crosses $e$ with the white site on the right or left, respectively. The definition \eqref{3.2} is well posed because the right side does not 
depend\footnote{More precisely, the values of the right side of \eqref{3.2} computed along two paths $C_{\eta\to\eta'}$ and $C_{\eta\to\eta'}'$ are the same if the loop obtained by concatenating 
$C_{\eta\to\eta'}$ with the path obtained by reversing the orientation of $C_{\eta\to\eta'}'$ is contractible. If $\Omega$ is a torus, then the two values may differ
by a quantity depending on the windings of such a loop.}
on the choice of the path $C_{\eta\to\eta'}$. 

The probability measure $\mathbb P^\lambda_{a,\Omega}$ depends on the parameters $t_1,t_2,t_3,\lambda$ and on the interaction $V$. Let us fix the latter once and for all. 
We are interested in identifying choices of $t_1,t_2,t_3,\lambda$ producing a non-trivial scaling limit as $a\to 0$ and/or $\Omega\nearrow\mathbb R^2$. However, contrary to the Ising 
case, the properties of the limiting distribution are extremely sensitive to the shape of $\Omega$, to the choice of its discretization and on the boundary conditions. 
Let us first consider the case that $\Omega=\Omega_L$ is a torus, centered at the origin, whose horizontal and vertical sides are both of length $L$.
In the limit $L\to\infty$, the 
expectation of the height function converges to a linear profile with  slope $\rho=\rho(t_1,t_2,t_3,\lambda)\in\mathbb R^2$: 
\begin{equation} \lim_{L\to\infty}\mathbb E^\lambda_{a,\Omega}(a h(\eta_x))=\rho\cdot x, \qquad \forall x\in\mathbb R^2\label{eq:3.3}\end{equation}
where, for $x\in \mathbb R^2$, $\eta_x$ is the face whose bottom vertex is black, of coordinate $[x]:=a\lfloor a^{-1}x\rfloor$. 
An alternative way of computing the function $\rho(t_1,t_2,t_3,\lambda)$ is via the Legendre transform of the free 
energy of the system with respect to a suitable `magnetic field' $B\in\mathbb R^2$: let $t_1(B)=t_1e^{-B_1}$, $t_2(B)=t_2e^{-B_1-B_2}$, $t_3(B)=t_3e^{-B_2}$ and 
\begin{equation} F(B):=\lim_{L\to\infty} L^{-2} \log Z^\lambda_{a,\Omega}(B),\end{equation}
with $Z^\lambda_{a,\Omega}(B)=\sum_{D\in\mathcal D}\Big(\prod_{e\in D} t_{r(e)}(B)\Big)e^{\lambda V(D)}$ the partition function with $B$-dependent weights, and define  
the \textit{surface tension} $\sigma: \mathbb R^2\to\mathbb R\cup\{+\infty\}$ as
\begin{equation} \sigma(s)=\sup_{B}\{s\cdot B-(B_1+B_2)/2-F(B)\}.\end{equation}
Then the average slope $\rho$ in \eqref{eq:3.3} is the unique minimizer of $\sigma$ w.r.t. $s$. 
If $\rho$ belongs to the region $\mathcal C$ where $\sigma$ is \textit{strictly convex} and twice differentiable, we also expect that the height fluctuations on top of the linear profile with slope $\rho$ 
is universally described by a Gaussian Free Field (GFF), in the sense that, for any $C^\infty$ compactly supported test function $\psi:\mathbb R^2\to\mathbb R$ 
such that $\int_{\mathbb R^2}\psi(x)dx=0$ and any $\alpha\in\mathbb R$, letting 
$h^a(\psi)=a^2\sum_{x\in\Omega_a} \psi(x)(h(\eta_x)-a^{-1}\rho\cdot x)$,  
\begin{equation}\lim_{a\to 0}\lim_{L\to\infty}
\mathbb E^\lambda_{a,\Omega}(e^{i\alpha h^a(\psi)})=e^{-\frac{\alpha^2}2\int_{\mathbb R^2} dx \int_{\mathbb R^2}dy\, \psi(x) \psi(y) G_\rho(x,y)}\label{eq:3.6}
\end{equation}
where $G_\rho$ is the Green's function, i.e., the inverse of $-\Delta_{\rho}:=-\sum_{i,j=1}^2\partial_i(\sigma_{ij}(\rho)\partial_j)$, with $\sigma_{ij}(\rho)$ the elements 
of the Hessian of $\sigma$ at $\rho$. 

We are now in the position of formulating a conjecture on the scaling limit of the dimer model in more general domains, involving the surface tension $\sigma$ introduced above and 
the region $\mathcal C$ of slopes where $\sigma$ is strictly convex and twice differentiable. Suppose, e.g., that $\Omega$ is an open, finite, simply connected region of the plane. 
Let $\bar h:\Omega \to \mathbb R$ be a continuous function that extends continuously to $\partial\Omega$, which is a `dimer limit shape', in the sense that it is the unique
minimizer of $\int_\Omega \sigma(\nabla h) dx$ with boundary condition $h\big|_{\partial\Omega}=
\bar h\big|_{\partial\Omega}$, and suppose that it has `no frozen regions', in the sense that $\nabla\bar h$ belongs to $\mathcal C$ for almost-every $x\in\Omega$. Then we expect that
there exists a sequence of discretizations $\Omega_a$ of $\Omega$ such that the average limiting height profile is exactly $\bar h$, i.e., $\lim_{a\to0}\mathbb E^\lambda_{a,\Omega}(ah(\eta_x))=\bar h(x)$, $\forall x\in \Omega$, and the scaling limit of the height fluctuations around $\bar h$ is a GFF, in the sense that, for any $C^\infty$ compactly supported test function $\psi:\Omega\to\mathbb R$, 
\begin{equation}\lim_{a\to 0}\mathbb E^\lambda_{a,\Omega}(e^{i\alpha h^a(\psi)})=e^{-\frac{\alpha^2}2\int_{\mathbb R^2} dx \int_{\mathbb R^2}dy\, \psi(x) \psi(y) 
G_{\bar h,\Omega}(x,y)}\label{eq:3.7}\end{equation}
where $G_{\bar h,\Omega}$ is the inverse of the operator $-\Delta_{\bar h}$ on $\Omega$ defined by \begin{equation}(-\Delta_{\bar h} f)(x):=
-\sum_{i,j=1}^2\partial_i(\sigma_{ij}(\bar h(x))\partial_j f(x)),\label{eq:3.8}\end{equation} with zero Dirichlet boundary conditions at 
$\partial\Omega$. The expected structure of the scaling limit is even richer than what emerges from the previous discussion: e.g., it turns out that the GFF nature of the 
height fluctuations is strictly (and subtly) related to the mesoscopic and macroscopic behavior of the dimer correlations, as it will become clearer from the discussion in the next two subsections. 
Moreover, the conjectured GFF nature of the height field comes together with complementary (and even harder-to-prove) predictions on the monomer and `vertex', or `electric', correlation functions, whose precise description, however, goes beyond the purpose of this review. 

\subsection{The non-interacting case}\label{sec:3.1}

As anticipated above, and in analogy with what we saw for the Ising model, at $\lambda=0$ the dimer model is exactly solvable \cite{Ka,TF}.
Let us review here a few aspects of the solution, and let us focus for simplicity on the case that $\Omega=\Omega_L$ is a square torus of side $L$, as described before \eqref{eq:3.3}. For any finite $a$ and $L$, the partition function can be expressed as the linear combination of the determinants of four variants of the so-called Kasteleyn matrix $K=K(t_1,t_2,t_3)$ (a complex adjacency matrix of $G_{\Omega_a}$), the four variants differing for the boundary conditions along the edges `winding up' over the torus, which can 
be periodic or anti-periodic in the horizontal and vertical directions. Moreover, the multipoint dimer correlations are (linear combinations of four) determinants of minors of $K^{-1}$. 
Starting from these explicit formulas, one can easily compute the limit of the dimer correlations as $L\to\infty$, thus finding, in particular, that, for any two edges $e,e'$, letting 
$r(e)\equiv r$, $r(e')\equiv r'$, $x(e)\equiv x$, $x(e')\equiv x'$: 
\begin{equation} \lim_{L\to\infty}\mathbb E^0_{a,\Omega}(\mathds 1_e\mathds 1_{e'})= K_{r} K_{r'} \det \begin{pmatrix} K^{-1}(x+av_{r}, x) & K^{-1} (x+av_{r}, x') \\
K^{-1}(x'+av_{r'}, x) & K^{-1}(x'+av_{r'}, x') \end{pmatrix}\end{equation}
where $K_r=i^{r-1}t_r$, $v_1=(0,0)$, $v_2=(-1,0)$, $v_3=(-1,-1)$, $v_4=(0,-1)$, and $K^{-1}$ is the inverse Kasteleyn matrix in the thermodynamic limit, which reads
\begin{equation} K^{-1}(x,y)=\int\limits_{[-\pi,\pi]^2} \frac{dk}{(2\pi)^2}\frac{e^{-ia^{-1}k\cdot(x-y)}}{\mu(k)},\label{eq:3.10}\end{equation}
with $\mu(k)=t_1+i t_2e^{ik_1}-t_3e^{i(k_1+k_2)}-ie^{ik_2}$ the `dispersion relation'. Using also the fact that $\lim_{L\to\infty}\mathbb E^0_{a,\Omega}(\mathds 1_e)= K_{r}
K^{-1}(x+av_{r},x)$ and $\lim_{L\to\infty}\mathbb E^0_{a,\Omega}(\mathds 1_{e'})= K_{r'}K^{-1}(x'+av_{r'},x')$, one finds that the truncated dimer-dimer correlation reads:
\begin{equation}\label{3.6} \lim_{L\to\infty}\mathbb E^0_{a,\Omega}(\mathds 1_e;\mathds 1_{e'})=-K_{r}K_{r'}K^{-1}(x+av_{r}, x')K^{-1}(x'+av_{r'},x), \end{equation}
whose large distance decay properties are dictated by those of $K^{-1}$. In turn, these depend on the singularity structure of $\mu(k)$: if $\mu$ has two simple zeros, 
denoted $p_+$ and $p_-$, a simple asymptotic computation shows that at large distances 
\begin{equation} K^{-1}(x,y)=\frac{a}{2\pi}\sum_{\omega=\pm} \omega\frac{e^{-ia^{-1}p_\omega\cdot (x-y)}}{\phi_\omega(x-y)}
+O((a/|x-y|)^2),\label{eq:3.12}\end{equation}
where $\phi_\omega(x)=\beta_\omega x_1-\alpha_\omega x_2$, with 
$\alpha_\omega=\partial_{k_1}\mu(p_\omega)$ and $\beta_\omega=\partial_{k_2}\mu(p^\omega)$. 
In view of \eqref{3.6}, 
\begin{eqnarray} && \lim_{L\to\infty}
\mathbb E^0_{a,\Omega}(\mathds 1_e;\mathds 1_{e'})=\frac{a^2}{4\pi^2} \sum_{\omega=\pm} \frac{K_{\omega,r}K_{\omega,r'}}{\big(\phi_\omega(x-x')\big)^2}\label{3.9}\\
&&\qquad +\frac{a^2}{4\pi^2} \sum_{\omega=\pm} \frac{K_{-\omega,r}K_{\omega,r'}}{\big|\phi_\omega(x-x')\big|^2}e^{ia^{-1}(p_\omega-p_{-\omega})\cdot (x-x')}+O((a/|x-x'|)^{-3}),\nonumber 
\end{eqnarray}
where $K_{\omega,r}=K_r e^{-ip_\omega\cdot v_r}$. Notice that both the first and the second term in the right side decay at large distances (compared to the lattice spacing) as 
$(a/|x-x'|)^2$, but the second term behaves differently from the first, because it wildly oscillates on the lattice scale. 

From these formulas, via \eqref{3.2}, one can compute the average height profile and the asymptotics of the height fluctuations around the average. In particular, using the expression 
for the dimer one-point function, we find that the two components of the average slope at $\lambda=0$, in the sense of \eqref{eq:3.3}, are $\rho_j=\sum_{e\in C_{\eta\to\eta+a\hat e_j}}\sigma_e (K_{r(e)}K^{-1}(x(e)+av_{r(e)}, x(e))-1/4)$, with $j=1,2$, for any face $\eta$ (here $\hat e_j$, is the unit coordinate vector in direction $j\in\{1,2\}$). 
Remarkably, $\rho$ belongs to the region $\mathcal C$ where the surface tension $\sigma$ is strictly convex and twice differentiable iff $\mu$ has two distinct zeros. 
In this case, by computing 
the asymptotics of the height fluctuations around the average height profile, we find, as expected, a GFF behavior: consider, e.g., four distinct points in the plane, $x_1,\ldots, x_4\in\mathbb R^2$;
using \eqref{3.2}, write the covariance of the height differences between the faces at $x_1,x_2$, and at $x_3,x_4$ as
\begin{equation} \begin{split} & \lim_{a\to 0}\lim_{L\to\infty}\mathbb E^0_{a,\Omega} (h(\eta_{x_1})-h(\eta_{x_2});h(\eta_{x_3})-h(\eta_{x_4}))\\ &\quad =\lim_{a\to 0}\sum_{\substack{e\in C_{\eta_{x_1}\to\eta_{x_2}}\\ e'\in 
C_{\eta_{x_3}\to\eta_{x_4}}}}\sigma_e\sigma_{e'}
\lim_{L\to\infty}\mathbb E^0_{a,\Omega} (\mathds 1_e;\mathds 1_{e'}),\end{split}\label{3.11}\end{equation}
and plug the asymptotic formula for the truncated dimer-dimer correlation \eqref{3.9} in the right side of this equation; using the independence of the right side of \eqref{3.11} from the choice of $C_{\eta_{x_1}\to\eta_{x_2}}, 
C_{\eta_{x_3}\to\eta_{x_4}}$, choose these lattice paths to be well separated: by doing so, one finds that both the remainder $O((a/|x-x'|)^3)$ and the wildly oscillating terms in \eqref{3.9} give subdominant contributions to the right side of \eqref{3.11}, in the $a\to 0$ limit; we are then left with the contribution from the first term in the right side of \eqref{3.9} and, using the remarkable fact that, for any $x\in \mathbb R^2$ and $j\in\{1,2\}$,
\begin{equation} \sum_{e\in C_{\eta_x\to \eta_{x+ae_j}}}\sigma_e K_{\omega,r(e)}=-i\omega\partial_j\phi_\omega(x), \label{eq:3.15}\end{equation}
we finally get 
\begin{equation}\label{eq:3.16} \begin{split}&\lim_{a\to 0}\lim_{L\to\infty}\mathbb E^0_{a,\Omega} (h(\eta_{x_1}-h(\eta_{x_2});h(\eta_{x_3})-h(\eta_{x_4}))\\
&=-\frac1{2\pi^2}\text{Re} \int_{\phi_+(x_1)}^{\phi_+(x_2)}dz\int_{\phi_+(x_3)}^{\phi_+(x_4)}dz' \frac1{(z-z')^2}\\ &=\frac1{2\pi^2}\text{Re} \log \frac{(\phi_+(x_4)-\phi_+(x_1))(\phi_+(x_3)-\phi_+(x_2))}
{(\phi_+(x_4)-\phi_+(x_2))(\phi_+(x_3)-\phi_+(x_1))}.\end{split}\end{equation}
Similar computations can be performed for higher moments of the height fluctuations, from which one finds that, for any $n>2$ and any $2n$-ple of distinct points $x_1,\ldots,x_{2n}$, $\lim_{a\to 0}\lim_{L\to\infty}
\mathbb E^0_{a,\Omega}$ $(h(\eta_{x_1})-h(\eta_{x_2});\cdots; h(\eta_{x_{2n-1}})-h(\eta_{x_{2n}}))=0$. As a corollary, one finds \eqref{eq:3.6} at $\lambda=0$, with $G_\rho(x,y)=-\frac1{2\pi^2}\log|\phi_+(x-y)|$. 
Similar results can be extended to the case of finite, simply connected, domains of arbitrary shape: in particular, the GFF behavior of the height field in the sense of \eqref{eq:3.7}-\eqref{eq:3.8} has been proved in \cite{Ke07,La21}. 

Note that, remarkably, the prefactor in front of the logarithm in the right side of \eqref{eq:3.16} (the `stiffness' of the GFF) is independent of the slope; equivalently, $\det\sigma_{ij}(\rho)\equiv\pi^2$, irrespective of $\rho$, provided $\rho\in\mathcal C$.
This is a \text{very} special property of the non-interacting model, related to the fact that the spectral curve is an algebraic Harnack curve \cite{KOS}, and it is not expected to be robust under the 
addition of interactions. More in general, one expects that the GFF behavior of the height fluctuation relies on a subtle relation between the `stiffness' coefficient of the GFF, equal to $\sqrt{\det\sigma_{ij}(\rho)}/\pi$, and the critical exponent associated with the oscillating part of the dimer-dimer correlation. Such a connection is a restatement, in the dimer context, of a deep universality relation predicted by Kadanoff \cite{Ka77} and Haldane \cite{Ha81} for vertex models and Luttinger liquids, based on Coulomb gas and bosonization methods. In the next section I will present a rigorous statement of the Kadanoff-Haldane relation for interacting dimer models and I will 
discuss its role in the proof of the GFF behavior of the height field. 

\subsection{Interacting dimer models}

Let us now consider interacting dimers, described by \eqref{3.1} with $\lambda\neq0$. In this case, the exact solvability of the model and its underlying determinant structure break down, and no 
thermodynamic or correlation function can be written explicitly, in closed form. 
This is the same as for the Ising model, but actually, compared with the Ising case, here things are even more subtle: while in the class of non-planar perturbations of the Ising model 
the scaling limit is expected to be described by the \textit{same} critical exponents as the nearest-neighbor model, the interacting dimer correlations are expected to display a
complex behavior, with different decay exponents associated with their oscillatory and non-oscillatory parts; in particular, the decay exponent associated with the 
oscillatory part of the two-point dimer correlation (i.e., the analogue of the second term in the right side of \eqref{3.9}) is expected to be \textit{anomalous}, i.e., to depend continuously and non-trivially on $\lambda$, and to be 
related by a simple, universal, relation to the stiffness coefficient of the height function. 

As an illustration of the non-trivial large distance behavior of the correlation functions of the interacting model, let me first state a result about the asymptotics of the two-point dimer correlation, which generalizes Eq.\eqref{3.9} to the case $\lambda\neq0$. Consider, again, the case that $\Omega=\Omega_L$ is a torus of side $L$ centered at the origin, and, given 
two edges $e,e'$, let $x(e)\equiv x$, $x(e')\equiv x'$, $r(e)\equiv r$, $r(e')\equiv r'$. Then the following holds: 

\begin{theorem}[\cite{GMT17,GMT20}] 
Let $t_1,t_2,t_3$ be such that $\mu(k)$ has two distinct non-degenerate zeros, $p_\pm$. 
Then there exist constants $C,\lambda_0>0$ and functions $K^\lambda_{\omega, r}$, $H^\lambda_{\o,r},$  $\alpha^\lambda_\omega$, $\beta^\lambda_\omega$, $p^\lambda_\omega$, $\nu(\lambda)$, 
analytic in $\lambda$ for $|\lambda|<\lambda_0$, for which, letting $\phi^\lambda_\omega(x)=\beta_\omega^\lambda x_1-\alpha^\lambda_\omega x_2$, 
  \begin{eqnarray} && \lim_{L\to\infty}\mathbb E^{\lambda}_{a,\Omega}(\mathds 1_{e};\mathds 1_{e'})=  \frac{a^2}{4\pi^2} \sum_{\omega=\pm} \frac{K^\lambda_{\omega, r}
    K^\lambda_{\omega, r'}
  }{(\phi^\lambda_\omega(x-x'))^2} \label{3.17}  \\
  &&\ +\frac{a^{2\nu(\lambda)}}{4\pi^2} \sum_\omega \frac{H^\lambda_{-\omega, r}
    H^\lambda_{\omega, r'} }{|\phi^\lambda_\omega(x-x')|^{2\nu(\lambda)}} e^{ia^{-1}
    (p^\lambda_\omega-p^\lambda_{-\omega})\cdot(x-x')} +O((a/|x-x'|)^{3-C|\lambda|}).\nonumber \end{eqnarray}
 Moreover, $K^0_{\omega, r}=H^0_{\omega,r}=K_{\omega,r}$, $\alpha^0_\omega=\partial_{k_1}\mu(p_\omega)$, $\beta^0_\omega=\partial_{k_2}\mu(p_\omega)$, $p^0_\omega=p_{\omega}$, $\nu(0)=1$, 
 \begin{equation} \overline{\alpha^\lambda_\omega}=-\alpha^\lambda_{-\omega},\quad \overline{\beta^\lambda_\omega}=-\beta^\lambda_{-\omega},\quad \overline{K^\lambda_{\omega, r}}=K^\lambda_{-\omega,r},\quad \overline{H^\lambda_{\omega, r}}=H^\lambda_{-\omega,r}, \quad p^\lambda_+ + p^\lambda_-=(\pi,\pi),\end{equation}
and, generically in the choice of the interaction $V$, $\nu(\lambda)$ depends non-trivially on $\lambda$, i.e., $\nu'(0)\neq 0$. 
\label{thm:3}\end{theorem}
The proof of the theorem provides a constructive algorithm for 
computing the coefficients of the convergent power series in $\lambda$ for $K^\lambda_{\omega,r}, H^\lambda_{\omega,r}$, etc., but does not provide  
closed formulas for any of them. By comparing \eqref{3.17} with \eqref{3.9}, it is apparent that the interaction modifies the scaling of the oscillatory part of the dimer-dimer correlation, which acquires the 
`anomalous' critical exponent $\nu(\lambda)$: this may be larger or smaller than $1$, depending on the sign of $\lambda$; therefore, depending on whether the 
dimer interaction is repulsive or attractive, the oscillatory term, in absolute value, may be dominant or subdominant at large distances w.r.t. the non-oscillatory term. 
Let us remark that 
an explicit computation \cite{GT19} shows that, generically, not only $\nu'(0)$ is different from zero, but it also depends explicitly upon the average slope $\rho_j=\rho_j(t_1,t_2,t_3,\lambda)=\sum_{e\in C_{\eta\to\eta+a \hat e_j}}
\sigma_e(\lim_{L\to\infty}\mathbb E^\lambda_{a,\Omega}(\mathds 1_e)-1/4)$.

Once that the sharp asymptotics for the two-point dimer correlation is know, we can compute the variance of height fluctuations, in analogy with \eqref{3.11} and following discussion. With the same notation and assumptions 
as in \eqref{3.11}, we write
\begin{equation} \begin{split} & \lim_{a\to 0}\lim_{L\to\infty}\mathbb E^\lambda_{a,\Omega} (h(\eta_{x_1})-h(\eta_{x_2});h(\eta_{x_3})-h(\eta_{x_4}))\\ &\quad =\lim_{a\to 0}\sum_{\substack{e\in C_{\eta_{x_1}\to\eta_{x_2}}\\ e'\in 
C_{\eta_{x_3}\to\eta_{x_4}}}}\sigma_e\sigma_{e'}
\lim_{L\to\infty}\mathbb E^\lambda_{a,\Omega} (\mathds 1_e;\mathds 1_{e'});\end{split}\label{3.18}\end{equation}
then we plug the asymptotics \eqref{3.17} in the right side of this equation, and by choosing the paths $C_{\eta_{x_1}\to\eta_{x_2}}, C_{\eta_{x_3}\to \eta_{x_4}}$ well separated, we find that the contributions to the variance
from the second and third terms in the right side of \eqref{3.17} vanish as $a\to 0$. So also in the interacting case we are left with the contribution to the variance from the non-oscillating term in \eqref{3.17}, 
which we need to evaluate in the $a\to 0$ limit. In order for the involved sums to converge to well-defined, path-independent, integrals, we need the analogue of \eqref{eq:3.15} to hold in the interacting case, too. This is very 
hard, if not impossible, to check directly, due to the fact that the coefficients of the convergent power series in $\lambda$ for $K^\lambda_{\omega,r}$ and $\phi^\lambda_\omega$ are defined by extremely 
complicated, and different, algorithms. Nevertheless, we succeeded in proving the validity of an interacting analogue of \eqref{eq:3.15}, by making use of latttice Ward Identities (WI), in combination with hidden, chiral, WI for a continuum reference model that, in an appropriate sense, describes the \textit{infrared fixed point} of the interacting dimer model (see next section for a few additional comments on the ideas of the proof): 

\begin{theorem}[\cite{GMT17b,GMT20}] \label{thm:4} Under the same assumptions as Theorem \ref{thm:3}, one has
\begin{equation}\label{eq:32xl}
\sum_{e\in C_{\eta\to\eta+a\hat e_j}} \sigma_e K^\lambda_{\omega,r(e)}=-i\omega\sqrt{\nu(\lambda)}\,\partial_j\phi^\lambda_\omega(x),
\end{equation}
where  $\nu(\lambda)$ is the same as in \eqref{3.17}. Consequently, 
\begin{equation}\label{eq:3.16} \begin{split}&\lim_{a\to 0}\lim_{L\to\infty}\mathbb E^\lambda_{a,\Omega} (h(\eta_{x_1}-h(\eta_{x_2});h(\eta_{x_3})-h(\eta_{x_4}))\\
&=\frac{\nu(\lambda)}{2\pi^2}\text{Re} \log \frac{(\phi_+(x_4)-\phi_+(x_1))(\phi_+(x_3)-\phi_+(x_2))}
{(\phi_+(x_4)-\phi_+(x_2))(\phi_+(x_3)-\phi_+(x_1))}.\end{split}\end{equation}
\end{theorem}

An elaboration of the proof also implies that, for any $n>2$ and any $2n$-ple of distinct points $x_1,\ldots,x_{2n}$, $\lim_{a\to0}\lim_{L\to\infty}  \mathbb E^\lambda_{a,\Omega}\big((h(\eta_{x_1})-h(\eta_{x_2}));\cdots;$ 
$(h(\eta_{x_{2n-1}})-h(\eta_{x_{2n}}))\big)=0$, 
from which the asymptotic GFF behavior of the height field, in the sense of \eqref{eq:3.6}, follows. The reader should not underestimate the fact that the proof of such GFF behavior comes with an exact computation of 
the stiffness coefficient of the GFF, which turns out to be the \textit{same} as the critical exponent $\nu(\lambda)$ of the dimer-dimer correlation. This is a universal relation among 
critical exponents, equivalent to those predicted by Kadanoff and Haldane in the closely related contexts of vertex, Ashkin-Teller, and Luttinger liquid models. In particular, it is equivalent to the 
identity $X_p=X_e/4$ \cite[Eq.(13a)]{Ka77} between the polarization critical exponent $X_p$ and the 
energy critical exponent $X_e$ of the Ashkin-Teller model, an elusive exact scaling relation that Kadanoff predicted on the basis of formal bosonization methods and Coulomb gas techniques. 
In this sense, our result is a rigorous confirmation of the predictions of bosonization in the context of interacting dimer models, and it is related to the notion of `weak universality' discussed in Baxter's book \cite{Ba}; 
see also \cite[Section 1]{GMT17b} and \cite{MaICM} for additional discussions on the notions of bosonization and weak universality in the contexts of dimer, vertex, Ashkin-Teller and quantum spin chain models. 

\section{Methods and ideas behind the proofs}\label{sec:4}

A common feature of the problems and results stated above is that they concern non-solvable 2D models in the vicinity of an exactly solvable reference model at its \textit{free Fermi point}: this is a way of saying that 
the reference nearest-neighbor Ising and dimer models are exactly solvable in terms of determinants of appropriate, explicit, matrices. 
As well known \cite{CCK,GeM,Sa}, this allows us to express the partition and generating function of correlations of the reference, solvable, models in terms of Gaussian Grassmann integrals: in particular, 
for any $a>0$ and finite $\Omega$, the partition function of the dimer model at $\lambda=0$ can be written as 
\begin{equation}Z^0_{a,\Omega}=\int D\phi e^{-(\phi^+,K\phi^-)},\label{4.1}\end{equation}
where $K$ is the Kasteleyn matrix, and $\phi=\{\phi^+_x,\phi^-_x\}_{x\in\Omega_a}\equiv(\phi^+,\phi^-)$ is a collection of Grassmann variables; the Ising partition function
can be written analogously, with $K$ replaced by a different, but still explicit, matrix, and $\phi$ replaced by a collection of $4|\Omega_a|$ `real' Grassmann variables. 
Similarly, the generating functions of the dimer or energy correlations, in the two cases of dimers and Ising, can be also written as Gaussian Grassmann integrals, 
from which one can easily get 
closed formulas for the corresponding multipoint dimer or energy correlations, and prove that they satisfy an exact fermionic Wick rule at the lattice level, i.e., that they have 
determinant, or Pfaffian, form. 
 
Another common feature of the dimer and Ising models discussed in this paper is that their interacting, non-solvable, versions can be formulated exactly, at finite $a$ and finite $\Omega$, in terms of non-Gaussian Grassmann integrals \cite{AGG2,GGM12,GMT20}. For instance, the partition function of the interacting dimer model can be written as follows (again, the one of the non-planar Ising model admits an analogous representation): 
\begin{equation} Z^\lambda_{a,\Omega}= Z^0_{a,\Omega} \int P(D\phi) e^{V(\phi)}, \label{4.2}\end{equation}
where $P(D\phi)= D\phi e^{-(\phi^+,K\phi^-)}/\int D\phi e^{-(\phi^+,K\phi^-)}$, and 
$V$ is a `potential' of strength $\lambda$, which can be written as the sum of monomials in $\phi$ of order $2$, $4$, $6$, etc., with kernels that are analytic in $\lambda$ in a 
small neighborhood of the origin and decay exponentially to zero at large distances, with rate proportional to the inverse lattice spacing. The term in $V$ that is quadratic in $\phi$ can be
isolated from the rest of the potential and combined with the Gaussian `measure' $P(D\phi)$; after this rearrangement, the potential contains a quartic term, plus higher order, 
subdominant, terms. In this sense, both the interacting Ising and dimer models take the form of Grassmannian $\phi^4_d$ models in dimension $d=2$, somewhat reminiscent of the 
$\phi^4_d$ models studied by the constructive Quantum Field Theory (QFT) community since the early 1970s \cite{GJ}. Note that \eqref{4.2} provides an explicit algorithm for computing 
all the coefficients of the perturbative series in $\lambda$ for the partition function (similar considerations hold for the free energy and generating function of correlations): it is enough to 
expand the exponential and compute term by term the expectation of $V^n$ w.r.t. the Gaussian measure $P(D\phi)$, which can be easily done in terms of the fermionic Wick rule. 
What makes things non trivial is the fact that the covariance of the reference Gaussian measure, which for dimers is the (finite volume analogue of the) inverse Kasteleyn matrix $K^{-1}$ in 
\eqref{eq:3.10}, at criticality decays algebraically to zero at large distances (criticality corresponds to the condition that $\mu(k)$ has two simple zeros, for dimers; and to the condition 
that $\beta$ is set at the inverse critical temperature, for Ising). This implies that the naive bounds one can easily derive on the coefficient of the perturbative series are non-uniform in 
$a$ and in the system size; in order to be able to control the thermodynamic and $a\to0$ limits, one needs to exhibit subtle cancellations, whose identification requires systematic, multiscale, resummations of the perturbation series, and which are very hard, if not impossible, to prove by direct inspection of the original series. 

The approach developed over the years to identify these cancellations in critical systems is based on the ideas of Wilsonian RG. More specifically, the constructive RG approach used in the proofs of 
Theorems \ref{thm:1} to \ref{thm:4} is the one developed by Benfatto, Gallavotti, Mastropietro and coworkers \cite{BG90, BGPS, BM05} and reviewed in, e.g., 
\cite{GeM, Mabook}; see also the more recent works \cite{AGG2}, \cite{GMT17}, and \cite{GMR21}, which review and provide a pedagogical introduction to this method in the specific 
contexts of non-planar Ising models, interacting dimer models, and fermionic $\phi^4_d$ theories with long range interactions, respectively. At a very coarse level, the idea is to 
compute \eqref{4.2} recursively, first integrating out the degrees of freedom at length scales $\ell_0 2^{-N}\simeq a$ (here $\ell_0$ is the length unit and $-N=\lfloor\log_2(a/\ell_0)\rfloor$), 
then those at length scales $\ell_0 2^{-N+1}, \ell_0 2^{-N+2}, \ldots,\ell_0 2^{-h+1}$, etc. After each integration step, we re-express the partition and generating functions in a form 
analogous to \eqref{4.2}, with $P(D\phi)$ replaced by a Gaussian measure with covariance supported on length scales $\gtrsim\ell_0 2^{-h}$ and $V(\phi)$ replaced by an effective 
interaction $V^{(h)}(\phi)$ that, up to a rescaling, has a form similar to the original $V(\phi)$, with modified coupling constants in front of the quadratic, quartic, sextic, etc., contributions. 
A dimensional power counting shows that the terms that tend to expand under iterations (and, therefore, to produce divergences in perturbation theory) are the quadratic and quartic ones, which tend to grow linearly and logarithmically in $2^{N-h}$, respectively. These are the terms to be monitored and carefully looked at, in order to identify the cancellations that, if present, allow one to defined a resummed, convergent, perturbation theory. 

Let me describe the procedure at a slightly more technical level, focusing, for illustrative purposes, on the dimer case with $\Omega=\Omega_L$ a torus of side $L$, and 
neglecting in the following discussion finite size effects, e.g., the difference between $K^{-1}$ and its finite-$L$ counterpart.
While the scheme described below has several similarities
with that used in the case of non-planar Ising models, there are also important differences (e.g., the presence for dimers of a non-trivial effective quartic coupling, denoted 
$\lambda_h$ in the following), which I will comment about below. 

In \eqref{4.2}, we first rewrite the covariance $K^{-1}(x,y)$ of the Gaussian measure as a superposition of exponentially decaying `propagators', each 
characterized by an exponential decay rate $\propto 2^h$, $h\le N$; that is, recalling \eqref{eq:3.10} and \eqref{eq:3.12}, we rewrite $K^{-1}(x,y)=\sum_{\omega=\pm}
\sum_{h\le N} e^{-ia^{-1}p_\omega\cdot(x-y)}g^{(h)}_\omega(x,y)$, with $g^{(h)}_\omega(x,y)\simeq 2^h g^{(0)}_\omega(2^hx, 2^hy)$, 
and $g_\omega^{(0)}$, $\omega=\pm$, two smooth functions, exponentially decaying to zero on scale $\ell_0$. 
Next, we rewrite the components of the random field $\phi$ with reference distribution $P(D\phi)$ as $\phi^\pm_x=\sum_{\omega=\pm}e^{\pm i a^{-1}p_\omega x}\phi^\pm_{\omega,x}$, 
and let $\phi_\omega=\phi_\omega^{(N)}+\phi_\omega^{(\le N-1)}$ (equality to be understood in distribution), with $\phi_\omega^{(N)}$ (resp. $\phi_\omega^{(\le N-1)}$) a Grassmann Gaussian field with covariance $g_\omega^{(N)}$
(resp. $g_\omega^{(\le N-1)}=\sum_{h\le N-1}g_\omega^{(h)}$); for brevity, we shall denote by $\phi^{(N)}$ the pair of fields $\{\phi^{(N)}_+,\phi^{(N)}_-\}$, and similarly for $\phi^{(\le N-1)}$. Correspondingly, we re-express the interacting partition function in \eqref{4.2} as follows (here $V^{(N)}$ is the same as $V$, thought of as a function of $\phi^{(N)}+\phi^{(\le N-1)}$ rather than of $\phi$):
\begin{equation} \begin{split}Z^\lambda_{a,\Omega}/Z^0_{a,\Omega}&= \int P_{\le N-1}(D\phi^{(\le N-1)})\int P_N(D\phi^{(N)}) e^{V^{(N)}(\phi^{(N)}+\phi^{(\le N-1)})}\\
&=e^{L^2 F_N} \int P_{\le N-1}(D\phi^{(\le N-1)})e^{V^{(N-1)}(\phi^{\le N-1})},\end{split} \label{4.3}\end{equation}
where $P_N(D\phi)$ and $P_{\le N-1}(D\phi)$ are the Grassmann Gaussian integrations with covariances $\int P_N(D\phi)\phi^-_{\omega,x}\phi^+_{\omega',y}=\delta_{\omega,\omega'}g_\omega^{(N)}(x,y)$ and $\int P_{\le N-1}(D\phi)\phi^-_{\omega,x}\phi^+_{\omega',y}=\delta_{\omega,\omega'}g_\omega^{(\le N-1)}(x,$ $y)$, respectively, and $L^2F_N+
V^{(N-1)}(\phi)=\log \int P_N(D\phi') e^{V(\phi'+\phi)}$, with $V^{(N-1)}(0)=0$; $F_N$ is a single-scale contribution to the free energy, and 
$V^{(N-1)}$ is called the effective potential on scale $2^{-N+1}$. Remarkably, $F_N$ and the kernels of $V^{(N-1)}$ are \textit{analytic} functions of $\lambda$, \textit{uniformly} in 
$a$ and $L$, thanks to the Grassmann nature of the theory (the key idea is that the $n$-th order term in perturbation theory can be expressed in determinant form, thanks to 
a smart interpolation identity due to Battle, Brydges, Federbush and Kennedy \cite{BaF,BF,BK}, and the determinants can be bounded in an optimal way, from the combinatorial point of view, thanks to the 
Gram-Hadamard inequality \cite{GeM}); moreover, the kernels of $V^{(N-1)}$ decay exponentially to zero at large distances, with exponential rate $\propto 2^{N-1}$. 

In the second line of \eqref{4.3}, we isolate the 
quadratic terms of $V^{(N-1)}(\phi)$ from the quartic or higher order terms, and we insert them in the reference Gaussian integration $P_{\le N-1}(D\phi)$, thus `dressing' it a little bit, the 
dressing corresponding to a small, $O(\lambda)$, change of the location of the zeros $p_\omega$ of $\mu(k)$, to an $O(\lambda)$ change of the `velocities'  
$\alpha_\omega=\partial_{k_1}\mu(p_\omega)$ and $\beta_\omega=\partial_{k_2}\mu(p_\omega)$, and to an overall rescaling by a multiplicative factor $Z_{N-1}=1+O(\lambda)$, 
which can be conveniently reabsorbed by rescaling the field $\phi$ by $\sqrt{Z_{N-1}}$. 
After the manipulation of these quadratic terms and this rescaling we are left with a modified effective interaction which includes a local quartic term, of the form 
$\lambda_{N-1}Z_{N-1}^2\int dx \phi^+_{+,x}\phi^-_{+,x}\phi^+_{-,x}\phi^-_{-,x}$, the constant $\lambda_{N-1}$ playing the role of the effective interaction strength on scale $2^{-N+1}$, 
plus a remainder, which is non local, or involves monomials in $\phi$ of higher order than four. 

We now iterate the procedure, and integrate out in the same fashion the fields on scales labelled by $N-1,N-2,\ldots, h+1$, so that, for any $h\le N$, we rewrite: 
\begin{equation} Z^\lambda_{a,\Omega}/Z^0_{a,\Omega}=e^{L^2\sum_{h'=h+1}^N F_{h'}} \int P_{\le h}(D\phi^{(\le h)})e^{V^{(h)}(\sqrt{Z_h}\phi^{\le h})},\label{4.4}\end{equation}
where, once again, the single-scale contributions to the free energy $F_{h'}$ and the kernels of the effective potential $V^{(h)}$ are analytic functions of $\lambda$, uniformly in $a,L$
(but, in general, non-uniformly in $N-h$). The constant $Z_h$ in the argument of the effective potential is the so-called wave-function renormalization, which plays the same role 
as the multiplicative factor $Z_{N-1}$ introduced above, after the integration of the first scale. Moreover, $V^{(h)}(\sqrt{Z_h}\phi)$ consists of: (i) 
quadratic terms, which can be combined with the reference Gaussian integration $P_{\le h}(D\phi^{(\le h)})$, 
thus leading to an additional, iteratively defined, dressing of the effective covariance; (ii) a local quartic term, of the form 
$\lambda_{h}Z_h^2\int dx \phi^+_{+,x}\phi^-_{+,x}\phi^+_{-,x}\phi^-_{-,x}$, with $\lambda_h$ playing the role of the effective interaction strength at length scales $\propto 2^{-h}$;
(iii) a remainder, including non-local interactions, or interactions of order higher than four in $\phi$, called the \textit{irrelevant} terms. 

As mentioned above, while well-defined at each scale, the procedure sketched above does not lead to bounds that, in general, are uniform in the number of iterations, $N-h$. Of course, in order to perform the scaling limit $a\to 0$ (which corresponds to the removal of the ultraviolet cutoff $N\to\infty$) and/or the thermodynamic limit $L\to\infty$ (which corresponds
to the removal of an infrared cutoff $h_L=\lfloor\log_2(\ell_0/L)\rfloor\to-\infty$), we need to prove that the construction is well defined uniformly in $N-h$. Remarkably, it turns out that 
the bounds on the kernels of the effective potential are uniform in the number of iterations iff $\lambda_h$ remains bounded and small, uniformly in the scale index: if this is the case, 
then all the irrelevant terms turn out to be bounded and small, too; in fact, they can be recasted in the form of uniformly convergent expansions in the \textit{effective} couplings 
$\{\lambda_{h'}\}_{h\le h'\le N}$. In other words, all the potential sources of divergences are resummed into the scale-dependent couplings $\{\lambda_{h'}\}$ and the problem 
of proving bounds on the free energy and correlation functions of the dimer model that are uniform in the scale index translates into that of controlling the boundedness of the
sequence of effective couplings. This is an enormous conceptual simplification, because $\lambda_h$ can be written as the solution to a finite difference equation, induced by the 
iterative integration procedure sketched above, known as the \textit{beta function} equation, of the form $\lambda_{h-1}=\lambda_h+\beta_h(\lambda_h,\ldots,\lambda_N)$, 
with $\beta_h(\lambda_h,\ldots,\lambda_N)=c_{2,h}\lambda_h^2+$ higher orders. A priori, the same general estimates leading to the aforementioned control on the irrelevant 
contributions to the effective potential tell us that $|\beta_h(\lambda_h,\ldots,\lambda_N)|\le C_0\epsilon_h^2$, with $\epsilon_h=\max_{h\le h'\le N}|\lambda_{h'}|$ and $C_0$ independent of $h$; therefore,
using the fact that $\lambda_N=O(\lambda)$, we find $|\lambda_h|\le C|\lambda|(1+|\lambda|(N-h))$ for some $h$-independent constant $C$; the point now is to look more closely to the beta function equation and try to identifying a structure guaranteeing that 
$\lambda_h$ behaves better than such a priori, general, bound. Explicit computations show that at second and third order $\beta_h(\lambda_h,\ldots,\lambda_N)$ is bounded by (const.)$2^{h-N}\epsilon_h^2$ 
and  (const.)$2^{h-N}\epsilon_h^3$, respectively. Analogous estimates at all orders would imply that $|\lambda_h|\le C|\lambda|$ uniformly in $h$, as desired. However, direct inspection of perturbation theory 
does not appear feasible, for bounding in a similar manner the general $n$-th order contribution to the beta function. 

The idea is to prove the desired cancellation via an indirect route: we introduce a reference model, which has the same beta function as the dimer model, asymptotically as $N-h\to\infty$, up to 
exponentiall small corrections, smaller than  $\epsilon_h^2 2^{\theta(h-N)}$, for some $\theta\in(0,1)$. This reference model plays the role of the `infrared fixed point' of our Grassmann formulation of the dimer model and 
is a close relative of the Luttinger model, an exactly solvable model of interacting fermions in one dimension, originally solved by rigorous bosonization techniques by Mattis and Lieb \cite{ML}. 
The reference model we use is a variation of the same model, formulated in the Grassmann functional integral setting, differening from the original Luttinger model `just' by the choice of the ultraviolet regulator (this 
apparently innocent modification may a priori have serious consequences, because exact integrability of the model requires a specific regularization scheme). 
Such a model displays additional symmetries as compared to the dimer model, most notably `local chiral gauge invariance'; i.e., the model is formally (up to corrections due to the ultraviolet regulator)
invariant under independent gauge transformations for the two chiral fields $\phi_\omega$, $\omega=\pm$. Chiral gauge invariance implies the validity of exact equations (`chiral Ward Identities') for the model's correlation 
functions; such equations include so-called `anomaly terms', i.e., terms that would naively be zero if one neglected the effects of the ultraviolet regulator, which, remarkably, can be computed explicitly, in closed form.
Combining such chiral Ward Identities (WI) with the so-called Schwinger-Dyson equation for the correlation functions, we are led to a closed formula for all the correlation functions of the moment. Such closed formulas imply, in particular, 
that the effective 
coupling strength $\lambda_{h,\text{ref}}$ of the reference model is uniformly close to its bare coupling; and, in turn, this implies the asymptotic vanishing of the beta function both for the reference and the dimer model. 
They also imply that the large distance asymptotic behavior of the dimer correlation functions are the same as those of appropriate correlations of the reference model: from this we derive the asymptotic formula 
\eqref{3.17} and prove Theorem \ref{thm:3}.

Not only that: the chiral WI imply exact identities (`scaling relations') relating different critical exponents, as well as critical exponents and the multiplicative prefactor in front of the density-density correlation. Such 
exact identities, if compared and combined with the exact lattice WI satisfied by the dimer correlation functions, imply analogous scaling relations for the dimer model;
in turn, the exact lattice WI of the dimer model are a consequence of the local conservation law for the number of incident dimers at each vertex. This is, 
at a rough level, the way in which we prove the 
identity \eqref{eq:32xl}, from which Theorem \ref{thm:4} and the GFF behavior of the height fluctuations follow. 

\medskip

Let me conclude this section by a brief discussion of how the previous strategy must be modified in order to prove Theorems \ref{thm:1} and \ref{thm:2} for non-planar Ising models. 
The general approach is the same: also the generating function of the energy correlations of these models can be expressed as  non-Gaussian Grassmann integral, similar to \eqref{4.2}. At the critical temperature, the 
covariance of the reference Gaussian integration decays algebraically to zero at large distances, and this implies that, in order to derive uniform bounds on the thermodynamic and correlation functions, 
we must appeal to a rigorous RG multiscale analysis. Therefore, also for Ising, we compute the non-Gaussian Grassmann functional integral in an iterative fashion, and we are led to 
the construction of a sequence of effective potentials $V^{(h)}$, in analogy with \eqref{4.4}. However, a crucial difference is in the counting of the `critical' degrees of freedom of the effective theory. In the Ising case, 
the effective potential $V^{(h)}$ can be written as the function of a Grassmann field $\phi$ with two components per site rather than four (remember: in the dimer case, there were four Grassmann variables per site, $\phi^+_{+,x}$, $\phi^-_{+,x}$, $\phi^+_{-,x}$, $\phi^-_{-,x}$, ; in the Ising case we have just two, $\phi_{+,x}$ and $\phi_{-,x}$, no $\pm$ label at exponent): this implies that the 
local quartic term in the effective potential, the one that was so hard to control in the dimer case, is automatically zero, because there is no non-vanishing quartic monomial that can be constructed (the Grassmann rule implies $\phi_{\omega,x}^2=0$). This makes the construction of the non-planar Ising theory in the full-plane limit easier than the one for interacting dimers (this also explain why Theorem \ref{thm:1}, which involves
non-planar Ising models in the full-plane limit, was proved already 10 years ago \cite{GGM12}). 

The problem now is the extension to finite domains with open, or cylindrical, boundary conditions: in fact, the presence of 
boundaries produce additional effective, scale-dependent, couplings, localized at the boundary, which are potentially logarithmically divergent, like the quartic effective coupling in the dimer setting. 
And, again, the proof that such additional boundary scale-dependent couplings remain bounded and small, uniformly in the scale index, requires to identify cancellations in their flow equation. In the Ising setting, these cancellations follow from an approximate image rule for the fermionic covariance at the boundary; once we identified this cancellation, we managed to extend the construction of the scaling limit of energy correlations to the 
cylindrical setting, thus proving Theorem \ref{thm:2}. Our proof is currently restricted to a specific cylindrical geometry, which we need in order to identify the required boundary cancellations, and in 
order to obtain optimal bounds on the fermionic Green's function in the vicinity of the boundary. However, the technique itself underlying the proof of the theorem seems robust and I expect that it can be adapted, in perspective, to domains of arbitrary shape (even more, I expect that it will be capable to understand scaling limits of models in the Luttinger liquid universality class, such as interacting dimers, in finite domains). See next section for additional comments on these perspectives. 

\medskip

Due to space constraints, I cannot enter in more detail than this into the proofs of the main theorems presented in this paper. The purpose of this section was just to convey the main ideas we used and to highlight the 
strategy and main difficulties to be overcome in the proofs. For additional details, I refer the reader to the original papers, \cite{AGG1,AGG2,GGM12,GMT17,GMT17b,GMT20}.

\section{Further results, perspectives and open problems}\label{sec:5}

Let me conclude this review with a brief, certainly partial, discussion of related results, perspectives and open problems, whose understanding would represent in my opinion 
a major advance in our understanding of the scaling limit of 
2D non-planar Ising models and interacting dimers models (as well as of related classes of non-integrable statistical mechanics models, such as Ashkin-Teller and vertex models). I will state explicitly 
only problems that are more directly connected with the results and methods reviewed in this paper. Of course, there are plenty of other challenging, extremely interesting, open problems, 
concerning, e.g., the scaling limit of critical interfaces \cite{BPW,CDHKS}, the limiting validity of Virasoro algebra for an appropriate class of `dressed' observables \cite{HJVK}, 
and the construction of the massive scaling limit in the magnetic field direction \cite{CGN,CJN} (for non-planar Ising models), or the scaling limit of vertex and monomer correlations
\cite{Du15}, the scaling limit of the cycle-rooted spanning forest associated with the dimer configuration via the Temperley bijection \cite[Sect.2.1.2]{GMT17b}, the validity of Cardy's 
formula \cite{Af,BCN,IPRH} and, more generally, 
the computation of the sub-leading corrections to the free energy \cite{CP} (for interacting dimers). 

\subsection{Non-planar Ising models}

Let us first consider the class of non-planar Ising models described above, in Section \ref{sec:2}. 
There are a few extensions of the results of Theorems \ref{thm:1} and \ref{thm:2} which appear to be feasible on the basis of 
relatively straightforward extensions of the techniques underlying their proofs. I refer, in particular, to the computation of boundary spin correlations and boundary energy correlations (and mixed boundary spin, boundary 
energy, bulk energy correlations): it should be easy to show, on the basis of a mild extension of the proof of Theorem \ref{thm:2}, that their scaling limit can be written as the Pfaffian of an explicit anti-symmetric matrix, 
whose elements (involving the two-point boundary spin-spin correlations) can be written in closed form, and exhibit the expected boundary critical exponents. This would complement the results of \cite{ADTW}, 
by computing explicitly the scaling limit for a wide class of non-planar perturbations of the Ising model, not restricted to ferromagnetic pair interactions.

On the other hand, extension of Theorem \ref{thm:2}, or of its expected analogue for boundary spin correlations, to domains of more general shapes than flat cylinders, appears to be much harder. 
Already in the $\lambda=0$ case, the construction of the scaling limit in domains of arbitrary shape remained elusive for several decades, and has been completed in the last ten years thanks to the use of the  
highly non-trivial methods of discrete holomorphicity, in the form developed, among others, by 
Smirnov, Chelkak, Hongler, and Izyurov \cite{CHI,CHI2,CS,HS}. It would be extremely interesting to extend this construction to the interacting, non-planar case. 

\textbf{Open problem 1.} \textit{Compute the scaling limit of the multipoint energy correlations of non-planar Ising models in domains $\Omega\subset \mathbb R^2$ of arbitrary shape, for different boundary conditions, say open, $+$ or $-$. As a corollary, prove conformal covariance of the limit.}

One possibility to attack this problem is to extend the strategy sketched in the previous section to more general domains (already the case of the rectangle is non trivial). 
There are two key technical points to be understood: (1) how can we obtain sufficient control on the fermionic Green's function in situations where it cannot be diagonalized explicitly? (By `control' here I mean:
define its multiscale decomposition, with optimal bounds on its asymptotic behavior in the bulk and close to the boundaries; moreover, derive a Gram representation for the single-scale Green's function, 
with optimal dimensional bounds on the $L^\infty$ norm of the Gram vectors); (2) how do we prove the required cancellations on the boundary, `marginal', scale-dependent couplings? 
I believe that the most serious technical issue is the first. A solution may come from an effective combination of multiscale methods with those of discrete holomorphicity, which may lead to 
sharp bounds on the speed of convergence to the scaling limit already at the level of the $\lambda=0$ theory. 

In connection with this problem, I cannot avoid mentioning an exciting recent development due to Duminil-Copin and collaborators \cite{DKKMO}, who proved rotational invariance for the scaling limit (whenever it exists) of 
a wide class of 2D critical models, including Potts, 6V and the random cluster model. The proof is based on completely different ideas than ours, and involve the coupling of different instances of these models on different 
isoradial graphs, characterized by different discrete rotational invariance properties, via a sequence of star-triangle transformations. 

\medskip

The first open problem, stated above, concerns energy correlations. Of course, analogous results for the spin correlations would also be extremely interesting. However, the 
study of the scaling limit of spin correlations is notoriously difficult, already in the full-plane limit, even in $\lambda=0$ case. The reason is that the spin observable is non-local in the Grassmann representation, 
and, already in the integrable case, their understanding requires the use of special, sophisticated techniques. In the full-plane limit at $\lambda=0$, one can use Szego's lemma to extract the asymptotics of the spin-spin correlations in special directions \cite{McCWbook}; or, alternatively, one can use a set of quadratic finite difference equations, discovered by McCoy, Perk and Wu \cite{McPeW}, whose scaling limit is the 
Painlev\'e III equation. Multi-point spin correlations, both in the full plane limit and in finite domains, were understood much more recently, thanks to other, complementary techniques, namely discrete holomorphicity  
applied on a two-sheet discrete Riemann surface, associated with the original graph which the model is define on, with cuts connecting the locations of the spin observables \cite{CHI,CHI2,Du11,Du15}. It is unclear whether these ideas can be extended, and in case how, to the interacting setting, $\lambda\neq0$. 

\textbf{Open problem 2.} \textit{Compute the scaling limit of the spin correlations of non-planar Ising models, first in the full plane, then in finite domains.}

Already the case of the two-point spin correlation in the full plane limit is highly non-trivial, and its understanding would represent a breakthrough in the field, with a potential big impact on other problems that, 
at a heuristic level, are studied by formal bosonization and Coulomb gas techniques. 

\medskip

Another interesting set of open problems is related to the computation of the sub-leading corrections to the critical free energy of non-planar Ising models, 
which are expected to display subtle universality properties \cite{CP,DFSZ}. 
Fix $\beta=\beta_c$, fix $\Omega$, and compute the free energy for $a$ small; it is expected that 
\begin{equation}
\log Z^\lambda_{\beta_c;a,\Omega}=a^{-2}|\Omega|f(\lambda)+a^{-1}|\partial\Omega|\tau(\lambda)+c_\Omega(a,\lambda), \end{equation}
with $f(\lambda)$ and $\tau(\lambda)$ independent of $\Omega$, and $c_\Omega(a,\lambda)$ of smaller order than $O(a^{-1})$. More precisely, it is expected that the behavior of this subleading term in the $a\to0$ depends upon the Euler characteristics $\chi$
of $\Omega$ (recall: $\chi=V-E+F$ with $V,E,F$ the number of vertices, edges, faces of any triangulation of $\Omega$; e.g., $\chi=0$ for $\Omega$ a torus or a cylinder, and $\chi=1$ for $\Omega$ a finite, simply connected domain). If $\chi\neq0$, it is expected that 
\begin{equation} \label{5.0}\lim_{a\to 0}\frac{c_\Omega(a,\lambda)}{\log(a^{-1}|\partial\Omega|)}=-\frac1{12} \chi, \end{equation}
while, if $\chi=0$, then 
\begin{equation}\lim_{a\to 0}c_\Omega(a,\lambda)=
c_\Omega^0\quad  \text{independent of $\lambda$}.\end{equation} 
E.g., if $\Omega=\Omega(\xi)$ is a torus with aspect ratio $\xi$, 
\begin{equation}c_\Omega^0=\log(\theta_2+\theta_3+\theta_4)-\frac13\log(4\theta_2\theta_3\theta_4),\label{5.1}\end{equation}
where $\theta_i=\theta_i(e^{-\pi \xi})$ are Jacobi theta functions. 

\textbf{Open problem 3.} \textit{Prove \eqref{5.0} and \eqref{5.1} with an explicit expression for $c^0_\Omega$ for non-planar Ising models.}

So far, the only known rigorous result related to this conjecture is a proof of `Cardy's formula' \cite{Af,BCN} for toroidal domains $\Omega=\Omega(\xi)$ with aspect ratio going to infinity: 
\begin{equation}\lim_{\xi\to\infty} \frac1{\xi}\lim_{a\to 0}c_{\Omega}(a,\lambda) = \frac{\pi}{12}. \label{eq:5.5}\end{equation}
Here, the right side is the $\xi\to\infty$ limit $\xi^{-1}$ times the right side of \eqref{5.1}. Eq.\eqref{eq:5.5} was proved in \cite{GM13}. 
I believe that the methods developed in \cite{AGG1,AGG2,GMT20} for controlling finite size effects within the rigorous RG scheme described in Section \ref{sec:4}
should be sufficient for proving \eqref{5.1} for the torus and the cylinder. Another story, which appears more challenging, is the case of $\chi\neq 0$. As far as I know, formula 
\eqref{5.0} is unproven even in the $\lambda=0$ case (with the exception of the rectangle, in which case it was proved in \cite{Hu17}). 

\subsection{Interacting dimer models}

Let us now consider the class of interacting dimer models discussed in Section \ref{sec:3}. 
All the problems stated in the previous subsection in the context of non-planar Ising models have their counterparts for dimers. Due to the underlying `Luttinger liquid' nature \cite{Ha81} 
of the 
scaling limit, and the presence of non-trivial, anomalous, exponents, I expect that their solution will be even more challenging than the one for the corresponding Ising's problems. 

\textbf{Open problem 4.} \textit{Prove the GFF nature of the scaling limit of the height fluctuations in arbitrary finite, simply connected, domains, in the sense of \eqref{eq:3.7}.}

In order to prove such a statement via the multiscale methods sketched in Section \ref{sec:4}, one will need to compute the dominant boundary corrections to the effective potentials, 
and, in particular, control the flow of the effective, marginal, boundary couplings (the dimer analogue of those discussed at the end of Section \ref{sec:4} for non-planar Ising). I expect that 
these boundary couplings will diverge exponentially in the limit of a large number of RG iterations, with a small, $\lambda$-dependent, exponent, playing the role of an anomalous 
boundary critical exponent. 

\textbf{Open problem 5.} \textit{Compute the asymptotic behavior, in the sense of \eqref{3.17}, for the two-point dimer correlation in a domain $\Omega$ with boundary, in the case in which at least one of the dimer 
observables is close to the boundary, and establish whether their oscillatory part exhibits an anomalous critical exponent $\nu_\partial(\lambda)$ \textit{different} from the bulk one $\nu(\lambda)$; in case, compute such boundary exponent.}

Other interesting directions and open problems involve generalizations of the type of dimer interactions.
For instance, rather than the class of interactions discussed in Section \ref{sec:3}, one could imagine to break planarity of the model, by adding 
non-planar, non-nearest-neighbor, edges to the graph, which may be occupied by `long' dimers with small probability. Under appropriate conditions on the geometry of these non-planar edges, e.g., 
if there exist lattice paths connecting faces microscopically close to any two points of the domain that never pass under the `bridges' formed by the non-planar edges\footnote{A concrete way of realizing this may be the following: consider a 2D periodic graph obtained by periodizing in two directions a planar fundamental cell $G_0$. Now make this non-planar, 
by adding in each fundamental cell a number of non-planar bonds, with the restriction that they should not pass over the `corridors' between different copies of $G_0$. Even though the graph is non-planar, the
height difference between faces in the corridors is well defined (use definition \eqref{3.2} with lattice paths passing only through the corridors).}
then it should still be possible to introduce a well-defined notion of height function. In such a situation, it would be interesting to test whether the GFF nature of the scaling limit of the height field persists, notwithstanding the loss of planarity of the model. 

\textbf{Open problem 6.} \textit{Same as Open Problem 4, for weakly non-planar dimer models.}

I expect that, at least in the case of periodic models defined on a torus of side $L$, in the limit $L\to\infty$, a generalization of the method of proof of Theorem \ref{thm:4} will allow us to 
prove convergence the convergence of a suitably defined height field to the massless GFF, in the sense of \eqref{eq:3.6}. Further extensions to different kind of dimer interactiosn appear 
more challenging. For instance, an extremely interesting problem that I propose here as my last Open Problem, is whether the GFF nature of the scaling limit of height fluctuations can be extended to the case of the interface between $+$ and $-$ 
phases in the 3D Ising model with `tilted' Dobrushin boundary conditions, at low enough temperatures. It is well known that the interface of the 3D Ising model with standard, flat, Dobrushin boundary conditions is rigid at low temperatures. This is not expected to be the case if the boundary conditions are assigned so that the interface has non-zero
average slope. The problem of understanding the nature of fluctuations of this interface, even if apparently very different from those considered in this review, 
has surprisingly strict connections with that of the scaling limit of the height fluctuations for interacting dimers \cite{CK01}: in fact, it is well known that the monotone height profiles 
of a tilted 3D Ising interface can be mapped exactly, in an invertible way, to those of the dimer model on the hexagonal lattice; moreover, under this mapping, 
the height distribution of the 3D Ising model at \textit{zero} temperature is the same as that of the standard, integrable, dimer model. From this exact correspondence, the GFF 
nature of the height fluctuations for the 3D Ising tilted interface readily follows. At positive temperatures, there is no known coupling between the height distribution of the 3D Ising tilted interface with that of a dimer model; however, it is tempting to guess that the effect induced by the temperature is qualitatively the same as that of a weak, effective, interaction among dimers. If this were the case, then the methods of Theorem \ref{thm:4} would provide a possible strategy for proving the existence of a `rough phase' for the 3D Ising model. 

\textbf{Open problem 8.} \textit{Prove that the fluctuations of the interface of the 3D Ising model with tilted Dobrushin boundary conditions at low temperatures converges in the scaling limit 
to a GFF.}

\medskip

\textbf{Acknowledgments.}
I would like to thank all my collaborators on the problems of universality and scaling limits in SM and quantum many body theory, and all the colleagues who shared their ideas with me in many inspiring discussions,
which were very influential for my research in this area. In particular, a special thank goes to Giovanni Antinucci, Rafael Greenblatt, Vieri Mastropietro and Fabio Toninelli, for the 
enjoyable and fruitful collaborations on Ising and dimer models, which led to the results reviewed in this essay.

This work has been supported by the European Research Council (ERC) under the European Union's Horizon 2020 research and innovation programme ERC CoG UniCoSM, grant agreement n.724939. I also acknowledge financial support from MIUR, PRIN 2017 project MaQuMA, PRIN201719 VMAST01.

\end{document}